\documentclass[12pt]{article}
\usepackage{amsmath}
\usepackage{multirow}
\usepackage{amsfonts}
\usepackage{amssymb,graphics,psfrag}
\usepackage{array,epsfig,multirow,stmaryrd,graphicx}
\usepackage{comment}
\usepackage{slashed}
\usepackage{hyperref}
\usepackage{tensor}
\usepackage{booktabs}

\def\hybrid{\topmargin -20pt    \oddsidemargin 0pt
        \headheight 0pt \headsep 0pt 
        \textwidth 6.25in      
        \textheight 9 in      
        \marginparwidth .875in
        \parskip 5pt plus 1pt
          \jot = 1.5ex
  }
\hybrid
\numberwithin{equation}{section}
\numberwithin{table}{section}

\newcommand{\beq}{\begin{equation}}
\newcommand{\eeq}{\end{equation}}
\newcommand{\be}{\begin{equation}}
\newcommand{\ee}{\end{equation}}
\newcommand{\bea}{\begin{eqnarray}}
\newcommand{\eea}{\end{eqnarray}}
\newcommand{\ben}{\begin{eqnarray*}}
\newcommand{\een}{\end{eqnarray*}}               
\newcommand{\ba}{\begin{aligned}}
\newcommand{\ea}{\end{aligned}}
\newcommand{\bt}{\begin{tabular}}
\newcommand{\et}{\end{tabular}}
\newcommand{\bc}{\begin{center}}
\newcommand{\ec}{\end{center}}

\newcommand{\cD}{\mathcal{D}}
\newcommand{\cL}{\mathcal{L}}

\newcommand{\cN}{\mathcal{N}}

\newcommand{\cH}{\mathcal{H}}

\newcommand{\cV}{\mathcal{V}}

\newcommand{\cM}{\mathcal M}

\DeclareMathOperator{\sign}{sign}

\newcommand{\nn}{\nonumber}

\newcommand{\cref}{{\bf [check ref]}}

\newcommand{\id}{\mathbf{1}}

\psfrag{n1}{$\nu_1$}
\psfrag{n2}{$\nu_2$}
\psfrag{n1'}{$\nu_1'$}
\psfrag{n2'}{$\nu_2'$}
\psfrag{n9}{$\nu_9$}
\psfrag{n10'}{$\nu_{10}'$}
\psfrag{t1}{${\nu}_1$}
\psfrag{t2}{${\nu}_2$}
\psfrag{t9}{${\nu}_9$}
\psfrag{t1'}{${\nu}_1'$}
\psfrag{t2'}{${\nu}_2'$}
\psfrag{t10'}{${\nu}_{10}'$}

\newcommand{\BB}{{\boldsymbol{B}}}
\newcommand{\BA}{{\boldsymbol{A}}}

\newcommand{\Bpsi}{{\boldsymbol{\psi}}}
\newcommand{\Bphi}{{\boldsymbol{\phi}}}
\newcommand{\Blambda}{{\boldsymbol{\lambda}}}

\def\blfootnote{\xdef\@thefnmark{}\@footnotetext}
\long\def\symbolfootnote[#1]#2{\begingroup%
\def\thefootnote{\fnsymbol{footnote}}\footnote[#1]{#2}\endgroup}

\begin{document}

\baselineskip=15pt

\begin{titlepage}
\begin{flushright}
\parbox[t]{1.8in}{\begin{flushright} MPP-2014-3 \end{flushright}}
\end{flushright}

\begin{center}

\vspace*{ 1.2cm}

{\large \bf  Self-Dual Tensors and  Partial Supersymmetry\\[.2cm]
                   Breaking in Five Dimensions}

\vskip 1.2cm

\begin{center}
 {Thomas W.~Grimm and Andreas Kapfer\ \footnote{grimm,\ kapfer@mpp.mpg.de}}
\end{center}
\vskip .2cm
\renewcommand{\thefootnote}{\arabic{footnote}}

{Max-Planck-Institut f\"ur Physik, \\
F\"ohringer Ring 6, 80805 Munich, Germany}

 \vspace*{1cm}

\end{center}

\vskip 0.2cm
 
 \begin{center} {\bf ABSTRACT } \end{center}

We study spontaneous supersymmetry breaking of 
five-dimensional supergravity theories from sixteen to eight 
supercharges in Minkowski vacua. This $\cN=4 \rightarrow \cN=2$ breaking 
is induced by Abelian gaugings that require the 
introduction of self-dual tensor fields accompanying the 
vectors in the gravity multiplet and vector multiplets. These tensor fields have first-order kinetic terms
and can become massive by a St\"uckelberg-like mechanism.
We identify the general class of $\cN=2$ vacua
and show how the $\cN=4$ spectrum splits into massless and massive 
$\cN=2$ multiplets.
In particular, we find a massive gravitino multiplet, containing two 
complex massive tensors, and a number of massive tensor multiplets
and hypermultiplets. We determine the resulting $\cN=2$
effective action for the massless multiplets obtained by 
integrating out massive fields. We show that the metric and Chern-Simons 
terms of the vectors are corrected at one-loop by massive tensors as well as spin-1/2 and spin-3/2
fermions. These contributions are independent of the supersymmetry-breaking 
scale and thus have to be included at arbitrarily low energies.

\vskip 0.4cm

\hfill {February, 2014}
\end{titlepage}

\tableofcontents

\newpage



\section{Introduction}

A systematic classification of supersymmetric vacua of supergravity theories in various dimensions 
has been a challenge since the first constructions of such theories. 
Supergravity theories with non-minimal supersymmetry can often admit 
Minkowski or anti-de Sitter ground states that preserve only a partial 
amount of supersymmetry. Finding such solutions is typically more involved 
than determining the fully supersymmetric solutions. 
For supergravity theories formulated in even space-time dimensions various breaking patterns 
have been investigated in detail. For example, the $\cN=2$ to $\cN=1$ breaking in four-dimensional 
supergravity theories has been investigated already in 
\cite{Cecotti:1984rk,Cecotti:1984wn,Cecotti:1985sf,Ferrara:1995gu,Antoniadis:1995vb,Fre:1996js,Kiritsis:1997ca,Andrianopoli:2002rm}. 
Recently, 
there has been a renewed interest in this direction \cite{Louis:2009xd,Louis:2010ui,Hansen:2013dda}, which was partially triggered by the application to flux 
compactifications of string theory \cite{Samtleben:2008pe}. The general analysis of \cite{Louis:2010ui} heavily employs 
the powerful techniques provided by the embedding tensor formalism \cite{deWit:2002vt,deWit:2005ub}.

The general study of partial supersymmetry breaking in 
odd-dimensional theories has attracted much less attention. 
Such theories, however, possess the interesting new possibility 
that the dynamics of some fields can arise from 
Chern-Simons-type couplings that are topological in nature. 
As was pointed out already for three-dimensional supergravity theories \cite{Hohm:2004rc}
such couplings can allow for special supersymmetry breaking patterns.
In this work we show that in five-dimensional supergravity theories 
with sixteen supercharges, denoted as $\cN=4$, Chern-Simons-type 
couplings for two-from tensor fields can yield interesting new supersymmetry 
breaking patterns to vacua preserving eight supercharges, denoted as $\cN=2$. 
Such tensor fields can have first-order kinetic terms
and become massive by a St\"uckelberg-like mechanism in which 
they eat a dynamical vector field \cite{Townsend:1983xs,Dall'Agata:2001vb,Schon:2006kz}. 
The degrees of freedom of such tensors are counted by 
realizing that they have zero degrees of freedom before eating the 
vector, but admit three degrees of freedom as
massive fields. Hence they should be distinguished from 
tensors with standard kinetic and mass terms. They have been named \textit{self-dual tensors} in \cite{Townsend:1983xs}.
In fact, such five-dimensional self-dual tensors arise, for example, as massive Kaluza-Klein modes 
of a six-dimensional self-dual tensor compactified on a circle \cite{Townsend:1983xs,Bonetti:2012fn}.
The mechanism rendering the tensor fields massive by eating a vector will be called 
\textit{tensorial Higgs mechanism} in the following.

In general,
the couplings of the five-dimensional self-dual tensors to the vector fields and the form of 
the first order kinetic terms are encoded by a constant anti-symmetric matrix $\xi_{MN}$, 
known as the embedding tensor \cite{Dall'Agata:2001vb,Schon:2006kz}. A non-trivial $\xi_{MN}$ also induces vector gaugings and a 
scalar potential. We analyze the conditions on $\xi_{MN}$ that yield partial supersymmetry breaking 
to an $\cN=2$ Minkowski vacuum. The massless and massive $\cN=2$ spectrum comprising 
fluctuations around this vacuum are then determined systematically. We particularly stress the
appearance of massive tensor fields and massive spin-1/2 and spin-3/2 fermions.
This allows us to derive the key features of the effective $\cN=2$ supergravity theory
arising for the massless fluctuations around the ground state. 

The $\cN=2$ effective action for the massless fields comprises of two parts. Firstly, 
there are the classical couplings inherited from the underlying $\cN=4$ theory. 
They are determined by truncating the original theory to the appropriately combined 
massless modes. At energy scales far below the supersymmetry breaking scale 
one might have expected that this determined already the complete $\cN=2$ theory.  
However, as we show in detail in this work, the massive tensor, spin-1/2 and spin-3/2 
modes have to actually be integrated out and induce non-trivial corrections. In 
fact, using the results of \cite{Bonetti:2013ela}, one infers that if these massive fields are 
charged under some vector field, they induce non-trivial one-loop corrections to the 
Chern-Simons terms for the vector. One-loop corrections to the Chern-Simons terms due to massive 
charged spin-1/2 fermions have been considered in \cite{Witten:1996qb,Morrison:1996xf}, but 
we stress here that in the $\cN=4$ to $\cN=2$ breaking both massive tensors and gravitini alter 
the result crucially. All these one-loop corrections are independent of the mass 
scale of these fields and therefore have to be taken into account in a consistent effective 
theory at scales well below the supersymmetry breaking scale.

Five-dimensional partial supersymmetry breaking can also occur
in string compactifications. For example, compactifications of 
Type IIB string theory on a certain class of five-dimensional 
compact spaces can yield five-dimensional 
$\cN=4$ theories that admit half-supersymmetric vacua as 
discussed, for example, in \cite{Gauntlett:2004zh,Cassani:2010uw,Gauntlett:2010vu,Liu:2010sa,Cassani:2010na,Bena:2010pr,Bah:2010cu,OColgain:2011ng}.
The generic analysis of these effective theories yields the existence also 
of  AdS vacua that are of particular interest for applications of the AdS/CFT duality. 
Our study of Minkowski vacua should be viewed as a crucial step in obtaining 
a full classification including also AdS vacua. We will see that already 
in the Minkowski vacua, arising due to the existence of tensor fields,
we can highlight interesting features of 
partial supersymmetry breaking in five dimensions. 
In particular, we show that the $\cN=2$ effective 
supergravity theory for the massless fluctuations around the $\cN=2$ vacuum
depends on the massive states of the underlying $\cN=4$ theory in an intriguing way.

The paper is organized as follows. In \autoref{N=4Generalities} we recall some 
general facts about $\cN=4$ supergravity theories. For an Abelian 
configuration with a non-trivial embedding tensor $\xi_{MN}$ we 
determine the conditions for the scalar potential to admit supersymmetric 
vacua. Supersymmetry breaking from $\cN=4$ to $\cN=2$ is 
studied in detail in \autoref{N=4toN2breaking}. We give a detailed account of 
the tensorial Higgs mechanism and the super-Higgs mechanism.
In \autoref{sec:effectiveaction} we then 
determine the $\cN=2$ spectrum and one-loop effective action. We argue that 
all massive multiplets induce one-loop corrections to the 
vector couplings of the theory that are independent of the 
supersymmetry breaking scale. In two supplementary appendices
we summarize our conventions and give details on the computation 
of fermion and scalar masses.

\section{Gauged $\cN =4$ supergravity in five dimensions} \label{N=4Generalities}

In this section we summarize some important facts about $\cN=4$ supergravity theories.
We first discuss the spectrum and action in \autoref{N=4Gen}. 
This will also allow us to state the conventions used throughout the paper. 
Further conventions and useful identities are supplemented in \autoref{App-Conventions}.
In \autoref{SusyCond} we study the vacua of an $\cN=4$ theory 
characterized by the components $\xi_{MN}$ of the embedding tensor
and comment on the possible amounts of supersymmetries preserved in 
the vacuum.

\subsection{Generalities} \label{N=4Gen}

Let us start reviewing the general properties of $\cN =4$ gauged supergravity in five 
dimensions following \cite{Dall'Agata:2001vb,Schon:2006kz}.\footnote{Let us stress again that in our conventions
five-dimensional $\cN =4$ supergravity theories have 16 supercharges and thus are half-maximal supergravities.} We first focus on 
ungauged Maxwell-Einstein supergravity that describes the coupling of 
$n$ vector multiplets and a gravity multiplet. Note that in the ungauged theory one can replace the
vector multiplets by dual tensor multiplets. 
The gravity multiplet consists of the fields
\begin{align} \label{grav_multiplet}
 (g_{\mu\nu}, \psi^i_\mu , A^{ij}_\mu , A^0 , \chi^i , \sigma )\, ,
\end{align}
with the metric $g_{\mu\nu}$, four spin-3/2 gravitini $\psi^i_\mu$, six vectors $(A^{ij}_\mu , A^0)$, four spin-1/2 fermions $\chi^i$
and one real scalar $\sigma$. The indices $i,j = 1,\dots ,4$ run over the fundamental representation of
the R-symmetry group $USp(4)$. We denote the symplectic form of $USp(4)$ by $\Omega$. It has the following properties
\begin{align}\label{e:properties_omega}
 \Omega_{ij} = - \Omega_{ji} \, , \qquad \Omega_{ij} = \Omega^{ij} \, , \qquad \Omega_{ij}\Omega^{jk} = - \delta^k_i \, .
\end{align}
Indices of $USp(4)$ are raised and lowered according to
\begin{align}\label{e:raising_lowering}
 V^i = \Omega^{ij} V_j \, , \qquad V_i = V^j \Omega_{ji} \, .
\end{align}
The \textbf{5} representation of $USp(4)$ is denoted by the double index $ij$ with the following properties
\begin{align}
 A^{ij}_\mu = - A^{ji}_\mu \, , \qquad A^{ij}_\mu \Omega_{ij} = 0 \, , \qquad (A^{ij}_\mu)^* = A_{\mu \, ij} \, .
\end{align}
Note that $USp(4)$ is the spin group of $SO(5)$. We will therefore often use the local isomorphism
$SO(5) \cong USp(4)$ to rewrite representations. We denote the indices of
the fundamental representation of $SO(5)$ by $m,n = 1, \dots , 5$. They are raised and lowered with the Kronecker delta $\delta_{mn}$.
We stress that all massless fermions in this work are symplectic Majorana
spinors. Further conventions and useful relations can be found in \autoref{App-Conventions}.
Moreover we will conveniently use the definition
\begin{align}
 \Sigma := e^{\sigma / \sqrt 3}\, ,
\end{align}
where $\sigma$ is the real scalar in the gravity multiplet \eqref{grav_multiplet}.

In addition to the gravity multiplet we have $n$ vector multiplets, labeled by the indices $a,b = 1, \dots , n$, which are raised
and lowered by the Kronecker delta $\delta_{ab}$. The field content is
\begin{align}
 (A_\mu^a , \lambda^{ia} , \phi^{ija})\, .
\end{align}
The $A_\mu^a$ denote vectors, $\lambda^{ia}$ spin-1/2 fermions and the $\phi^{ija}$ are scalars in the \textbf{5} of $USp(4)$.

Collecting all scalars of the theory they span the coset manifold 
\begin{align} \label{coset_def}
 \cM = \cM_{5,n} \times SO(1,1) \, , \qquad \quad \cM_{5,n} = \frac{SO(5,n)}{SO(5)\times SO(n)}\, ,
\end{align}
where the coset $\cM_{5,n}$ is parametrized by the scalars $\phi^{ija}$ in the vector multiplets, while
the $SO(1,1)$ factor is described by the scalar $\sigma$ in the gravity multiplet. Thus the global symmetry group of the theory
is $SO(5,n)\times SO(1,1)$.
We also introduce $SO(5,n)$ indices $M,N= 1,\dots 5+n$, which
are raised and lowered with the $SO(5,n)$ metric $(\eta_{MN}) = \textrm{diag}(-1,-1,-1,-1,-1,+1,\dots ,+1)$.
The generators $t_{MN}$ of $SO(5,n)$ and $t_0$ of $SO(1,1)$ are given by
\footnote{All antisymmetrizations in this paper include a factor of $1/n!\,$.}
\begin{align}
 \tensor{t}{_M_N_\, _P^Q} = 2\delta_{[M}^Q \eta_{N]P} \, , \qquad \tensor{t}{_0_\, _M^N} = - \frac{1}{2}\delta_M^N \, ,
 \qquad \tensor{t}{_M_N_\, _0^0} = 0 \, , \qquad \tensor{t}{_0_\, _0^0} = 1 \, .
\end{align}
To couple the vector multiplets to the gravity multiplet we note that the vectors of both kind of multiplets transform 
as a singlet $A^0$ and the fundamental
representation $A^M$ of $SO(5,n)$:
\begin{align}
 (A^0 , A^{ij} , A^n ) \rightarrow (A^0 , A^{M})
\end{align}
with $SO(1,1)$ charges $-1$ and $1/2$ for $A^0$ and $A^M$, respectively.

The coset space $\cM_{5,n}$ is most conveniently described by the coset representatives
$\cV=(\tensor{\cV}{_M^m} , \tensor{\cV}{_M^a})$, where $m = 1, \dots , 5$ is the $SO(5)$ index, while $a = 1, \dots n$ is the $SO(n)$ index. 
Global $SO(5,n)$ transformations on $\cV$ act from the left, while local $SO(5)\times SO(n)$ transformations act from the right.
We stress that
\begin{align}\label{e:SO(5,N)_index_raising}
 \tensor{\cV}{_M^a} = \eta_{MN}\cV^{Na} \, , \qquad \tensor{\cV}{_M^m} = - \eta_{MN} \cV^{Nm} \, .
\end{align}
Note that since
$(\tensor{\cV}{_M^m} , \tensor{\cV}{_M^a})\in SO(5,n)$ we have
\begin{align} \label{eta_viaV}
 \eta_{MN} = - \tensor{\cV}{_M^m} \cV_{N m} + \tensor{\cV}{_M^a} \cV_{N a} \, .
\end{align}
Let us also introduce a non-constant positive definite metric on the coset
\begin{align} \label{M_viaV}
 M_{MN} = \tensor{\cV}{_M^m} \cV_{N m} + \tensor{\cV}{_M^a} \cV_{N a} \, .
\end{align}
The inverse of $M_{MN}$ is given by $M^{MN}$.
We can also make use of the local isomorphism $SO(5) \cong USp(4)$ to define
\begin{align}\label{e:local_iso}
 \tensor{\cV}{_M^i^j} := \frac{1}{2}\tensor{\cV}{_M^m} \tensor{\Gamma}{_m^i^j} \,
 ,\qquad \tensor{\cV}{_M^m} = \frac{1}{2}\tensor{\cV}{_M^i^j} \tensor{\Gamma}{^m_i_j}\, ,
\end{align}
where $\tensor{\cV}{_M^i^j}$ transforms in the \textbf{5} of $USp(4)$ and $\tensor{\Gamma}{^m_i_j}$ are the components of the 
gamma-matrix $\Gamma^m$ of $SO(5)$.

There are now different possibilities to gauge some of the global symmetries. The various gaugings can be described in terms of the
embedding tensors $f_{MNP}$, $\xi_{MN}$ and $\xi_M$ that are totally antisymmetric in their indices. They specify a covariant derivative \footnote{Note that 
one can explicitly include a gauge coupling constant $g$ whenever an embedding tensor appears. For convenience we take $g=1$ in the following.}
\begin{align} \label{gen_cov_der}
 D_\mu = \nabla_\mu -  A_\mu^M \tensor{f}{_M^N^P} t_{NP} -  A_\mu^0 \xi^{NP} t_{NP} -  A^M_\mu \xi^N t_{MN} -  A_\mu^M \xi_M t_{0}\, .
\end{align}
We note that in the ungauged theory the embedding 
tensors are spurionic objects transforming under the global symmetry group.
As soon as we fix a value for the tensor components, the global symmetry group is broken down to a subgroup.
In this paper we will focus solely on the gauging $\xi_{MN}$, since the calculations simplify considerably.
Furthermore, we will find that a non-vanishing $\xi_{MN}$ is essential in the tensorial Higgs mechanism and interesting parts 
of the structure arising with general gaugings are already present in the case of $f_{MNP}=\xi_M = 0$. 
Therefore, we will from now on set
\begin{align} \label{vanishembedding}
 f_{MNP}=\xi_M = 0 \, ,
\end{align}
such that the covariant derivative \eqref{gen_cov_der} simplifies to
\begin{align}
 D_\mu = \nabla_\mu - A_\mu^0 \xi^{NP} t_{NP} \, .
\end{align}
We note that in the case of $f_{MNP}=\xi_M = 0$ there are no 
further constraints on $\xi_{MN}$ except of antisymmetry. 
It is important to remark that such a nontrivial gauging has the effect, that we are forced to dualize
some of the vector fields $A_\mu^M$ into two-forms $B_{\mu\nu\, M}$. Therefore, in order to write down the most general gauged
supergravity with $f_{MNP}=\xi_M = 0$, we have to consider 
an action where both $A_\mu^M$ and $B_{\mu\nu\, M}$ are present.
In this formulation the tensor fields $B_{\mu\nu\, M}$ carry no on-shell degrees of freedom,
they can, however, become massive by eating a dynamical vector and acquire three degrees of 
freedom. We will discuss this further in \autoref{tensorialHiggs}.

The bosonic Lagrangian of this $\cN=4$ supergravity theory reads \cite{Dall'Agata:2001vb,Schon:2006kz}
\begin{align} \label{bos_N=4action}
 e^{-1}\cL_{\textrm{bos}}=&
 -\frac{1}{2}R - \frac{1}{4}\Sigma^2 M_{MN} \cH^M_{\mu\nu} \cH^{\mu\nu\, N} - \frac{1}{4}\Sigma^{-4}F_{\mu\nu}^0 F^{\mu\nu\, 0}\nn \\
 &-\frac{3}{2}\Sigma^{-2}(\nabla_\mu \Sigma)^2 
  + \frac{1}{16}(D_\mu M_{MN})(D^\mu M^{MN})\nn \\
 &+\frac{1}{16\sqrt 2}\epsilon^{\mu\nu\rho\lambda\sigma} \xi^{MN}B_{\mu\nu\, M} ( D_\rho B_{\lambda\sigma\, N}
 +8 \eta_{NP}A^P_\rho \partial_\lambda A^0_\sigma  ) \nn \\
 & - \frac{1}{\sqrt 2}\epsilon^{\mu\nu\rho\lambda\sigma}\eta_{MN} A_\mu^0 \partial_\nu A^M_\rho \partial_\lambda A^N_\sigma \nn \\
 & - \frac{1}{16} \xi_{MN}\xi_{PQ} \Sigma^4 (M^{MP}M^{NQ}- \eta^{MP}\eta^{NQ})\, .
\end{align}
In this expression $R$ denotes the Ricci scalar and we define
\begin{align} \label{def-cH}
 \cH^M_{\mu\nu} := F^M_{\mu\nu} - 2 \tensor{\xi}{_N^M}A_{[\mu}^0 A_{\nu]}^N + \frac{1}{2} \xi^{MN} B_{\mu\nu\, N}\, ,
\end{align}
where
\begin{align}
 F^M_{\mu\nu} := \partial_\mu A^M_\nu - \partial_\nu A^M_\mu \, , \qquad F^0_{\mu\nu} := \partial_\mu A^0_\nu - \partial_\nu A^0_\mu \, .
\end{align}
The vectors are subject to gauge transformations with scalar parameters $(\Lambda^0 , \Lambda^M)$, while the tensors transform under
standard two-form gauge transformations with one-form parameters $\Xi_{\mu\, M}$. The variation of a vector under these
transformations will play a prominent role in the work of this paper, since it allows to implement the tensorial Higgs mechanism. It reads
for our choice of gaugings
\begin{align} \label{general_gaugetransform}
 \delta A^0_\mu = \nabla_\mu \Lambda^0 \, , \qquad \delta A^M_\mu = \nabla_\mu \Lambda^M - A^0_\mu \tensor{\xi}{_N^M} \Lambda^N
 -\frac{1}{2} \xi^{MN} \Xi_{\mu\, N}\, .
\end{align}

Next turn to the fermionic Lagrangian. For the purpose of this work it 
will be sufficient to only recall the kinetic terms, the mass terms as well as further terms quadratic in the fields.
A more complete discussion can be found in \cite{Dall'Agata:2001vb,Schon:2006kz}.
The quadratic fermionic terms of interest read
\begin{align} \label{fermionic-quadratic}
 e^{-1}\cL_{\textrm{ferm}}=&
 -\frac{1}{2} \bar \psi^i_\mu \gamma^{\mu\nu\rho}\cD_\nu \psi_{\rho\, i}
 -\frac{1}{2}\bar\chi^i \slashed\cD \chi_i -\frac{1}{2}\bar\lambda^{ia} \slashed\cD \lambda^a_i \\
 &+\frac{\sqrt 6 i }{4} A_{1\, ij} \bar \psi^i_\mu \gamma^{\mu\nu} \psi^j_\nu + 
 \Big(-\frac{1}{2\sqrt 2 }\Sigma^2 \xi_{ab} \Omega_{ij} + \frac{\sqrt 6}{4}A_{1\, ij} \delta_{ab}\Big ) i  \bar\lambda^{ia}\lambda^{jb} \nn \\
& - \frac{5\sqrt 6 i}{12}  A_{1\, ij} \bar\chi^i \chi^j
 -  \sqrt 2  A_{2\, aij} \bar\psi^i_\mu \gamma^\mu \lambda^{ja} - \sqrt 2  A_{1\, ij} \bar\psi^i_\mu \gamma^\mu \chi^j
 - \frac{4 \sqrt 6 i }{3} A_{2\, aij} \bar \chi^i \lambda^{ja} \, , \nn
\end{align}
where
\begin{align}\label{e:shift_matrices}
 A_1^{ij} = - \frac{\sqrt 3}{3} \Sigma^2  \Omega_{kl} \tensor{\cV}{_M^i^k} \tensor{\cV}{_N^j^l} \xi^{MN} \, , \qquad
 A_2^{aij} = - \frac{1}{2} \Sigma^2 \tensor{\cV}{_M^a} \tensor{\cV}{_N^i^j} \xi^{MN} \, . 
\end{align}
The first line in \eqref{fermionic-quadratic} are the kinetic terms of 
the fermions, while the remaining two lines summarize their mass terms.
The covariant derivatives are given by
\begin{align}
&\cD_\mu \psi_{i\, \nu} = \nabla_\mu \psi_{\nu\, i} + \frac{\sqrt 3 }{\Sigma^2} A_\mu^0 \tensor{A}{_1_\, _i^j} \psi_{\nu\, j}+ \dots \, , \\
&\cD_\mu \chi_i = \nabla_\mu \chi_i + \frac{\sqrt 3 }{\Sigma^2} A_\mu^0 \tensor{A}{_1_\, _i^j} \chi_j + \dots \, , \nn \\
 &\cD_\mu \lambda^a_i = \nabla_\mu \lambda_i^a + \frac{\sqrt 3 }{\Sigma^2} A_\mu^0 \tensor{A}{_1_\, _i^j} \lambda_j^a
 +  A_\mu^0 \tensor{\xi}{^a_b}\lambda_i^b + \dots \, , \nn
\end{align}
where the dots indicate couplings to scalars, which are of no importance in this work.
In order to simplify expressions like \eqref{fermionic-quadratic} we introduce some convenient notation. 
We denote contractions of the embedding tensor
with the coset representatives by
\begin{align}\label{e:dressed_gaugings}
 \xi^{mn} := \tensor{\cV}{_M^m} \xi^{MN} \tensor{\cV}{_N^n}\, , \qquad \xi^{ab} := \tensor{\cV}{_M^a} \xi^{MN} \tensor{\cV}{_N^b} \, , \qquad
 \xi^{am} := \tensor{\cV}{_M^a} \xi^{MN} \tensor{\cV}{_N^m} \, .
\end{align}
Thus these quantities are field-dependent and only become constant in the vacuum. Note that in the definition of
$\xi^{mn}$ and $\xi^{am}$ the positions of the $SO(5,n)$-indices $M,N$ are crucial because of \eqref{e:SO(5,N)_index_raising}.
The local isomorphism of $SO(5)$ and $USp(4)$ of \eqref{e:local_iso} establishes relations between
\eqref{e:shift_matrices} and \eqref{e:dressed_gaugings}. More precisely, one finds using \eqref{tensor_correspondence}
\begin{align} \label{ferm_mass}
 \xi^{mn} &= - \frac{\sqrt 3}{2 \Sigma^2}A_1^{ij}\tensor{\Gamma}{^m^n_i_j} \, , & \qquad 
 A_1^{ij} &= - \frac{1}{4 \sqrt 3}\Sigma^2 \xi^{mn} \tensor{\Gamma}{_m_n^i^j}\, , & \\
 \xi^{am} &= - \frac{1}{\Sigma^2} A_2^{aij} \tensor{\Gamma}{^m_i_j} \, , & \qquad
 A_2^{aij} &= - \frac{1}{4} \Sigma^2 \xi^{am} \tensor{\Gamma}{_m^i^j}\, , & \nn
\end{align}
where $\tensor{\Gamma}{_m_n^i^j}$ are the components of 
the gamma-matrix product $\Gamma_{mn} = \Gamma_{[m} \Gamma_{n]}$.

\subsection{Vacuum conditions and supersymmetry breaking} \label{SusyCond}

In this subsection we formulate the conditions for finding 
vacua of the gauged $\cN =4$ supergravity theory introduced above.
We restrict our considerations to the case for which  
$\xi_{MN}$ is the only non-vanishing embedding tensor such that \eqref{vanishembedding} holds.
The scalar potential $V$ found in \eqref{bos_N=4action} is then given by \cite{Dall'Agata:2001vb, Schon:2006kz}
\begin{align}\label{scalar_potential}
 V =  \frac{1}{16} \xi_{MN}\xi_{PQ} \Sigma^4 \big (M^{MP}M^{NQ} - \eta^{MP}\eta^{NQ} \big ) \, .
\end{align}
Inserting the explicit expressions \eqref{eta_viaV} and \eqref{M_viaV} of $\eta_{MN}$ and $M_{MN}$ in terms of the coset representatives gives
\begin{align}
\label{e:potential_1}
 V = \frac{1}{4}\xi^{MN}\xi^{PQ}\Sigma^4 \tensor{\cV}{_M^a} \tensor{\cV}{_P_a} \tensor{\cV}{_N^m} \tensor{\cV}{_Q_m} 
 &= \frac{1}{4}\Sigma^4 \xi^{am}  \xi_{am}  \, ,
\end{align}
where we have inserted \eqref{e:dressed_gaugings}.
Using the fact that $a$ and $m$ indices are raised by the 
Kronecker delta, this implies that the scalar potential is a sum of positive semi-definite terms.

Determining the minima of this potential is now trivial. The derivative with respect to $\Sigma$ yields
\begin{align}
 \frac{\partial V}{\partial \Sigma} =  \Sigma^3 \xi^{am}  \xi_{am} \overset{!}{=} 0\, .
\end{align}
Since the left-hand-side of this equation is a sum of non-negative quadratic terms, the solution simply reads\footnote{Let
us stress once more that $\xi^{am}$ is a field-dependent quantity.}
\begin{align}
\label{e:vacuum_condition}
  \xi^{am} \overset{!}{=} 0 \qquad\forall a,m \, .
\end{align}
The potential at this point is
\begin{align}\label{e:pot_at_minimum}
 V\big\vert_{\xi^{am} = 0 } = 0 \, .
\end{align}
The remaining derivatives with respect to the scalars in the vector multiplets are trivially 
vanishing since the potential is quadratic in the $\xi^{am}$. 

In summary, for vanishing embedding tensors $f_{MNP}$, $\xi_M$ the vacua are characterized by the condition
$\xi^{am} = 0$ for all $a,m$. Due to \eqref{e:pot_at_minimum} all such vacua are necessarily Minkowskian. 
Furthermore, we note that due to the vacuum condition the fermion shift matrix $A_2^{aij}$ given in \eqref{ferm_mass} trivially vanishes.

Let us next discuss the amount of supersymmetry preserved by these vacua.
It can be determined by analyzing the variations of the fermions under supersymmetry
transformations, which can be found in \cite{Dall'Agata:2001vb}. Evaluated in the vacuum \eqref{e:vacuum_condition} they read
\begin{align}
 \delta \psi_\mu^i =  - \frac{i}{\sqrt 6} \tensor{A}{_1_\, _i^j} \gamma_\mu \epsilon_j \, , \qquad
 \delta \chi_i = -\sqrt 2  \tensor{A}{_1_\, _i^j} \epsilon_j \, , \qquad
 \delta \lambda^a_i = 0 \, ,
\end{align}
where $\epsilon_i = \epsilon_i (x)$ is the supersymmetry parameter. The automatic vanishing of
$\delta \lambda^a_i$ is a direct consequence of the fact that $A_2^{aij} = 0$ in the vacuum.
It is convenient to decompose the supersymmetry parameter $\epsilon_i$ into a Killing spinor $\eta$ and a 
spacetime independent $USp(4)$ vector $q_i$
\begin{align}
 \epsilon_i = q_i \eta \, .
\end{align}
Each conserved supersymmetry corresponds to a zero eigenvalue of the matrix $\tensor{A}{_1_\, _i^j}$. 
Therefore we now need to face 
the task of finding the general form of the eigenvalues 
of $\tensor{A}{_1_\, _i^j}$.

The eigenvalues 
of $\tensor{A}{_1_\, _i^j}$ were already determined in \cite{Cassani:2012wc} and we will recall their analysis in the following. 
To begin with one considers the decomposition of the
$\cN=4$ R-symmetry group $USp(4)$ under the choice of a special direction.
Recall that there is the local isomorphism $USp(4) \cong SO(5)$. One then defines an $SO(5)$ vector
\begin{align}
 \tilde X^m := \varepsilon^{mnpqr}\xi_{np}\xi_{qr}\, ,
\end{align}
which specifies a preferred direction in $\mathbb{R}^5$ and therefore encodes the breaking
\begin{align} \label{group-split}
 SO(5) \rightarrow SO(4) \cong SU(2)_+ \times SU(2)_-\, .
\end{align}
Here $\varepsilon^{mnpqr}$ denotes the usual five-dimensional Levi-Civita symbol with
$\varepsilon^{12345}= 1$.
One also defines a unit vector for $\tilde X \neq 0$
\begin{align}
 X^m := \tilde X^m / \lvert \tilde X \rvert
\end{align}
with
\begin{align}
 \lvert \tilde X \rvert = \sqrt{\tilde X^m \tilde X_m}= \sqrt{8(\xi^{mn}\xi_{mn})^2 -16\xi^{mn}\xi_{np}\xi^{pq}\xi_{qm}} \, .
\end{align}
Via this construction, the four eigenvalues of the hermitian matrix 
$i\tensor{A}{_1_\, _i^j}$ are given by $\pm a_{1\pm}$ , where \cite{Cassani:2012wc}
\begin{align}
\label{e:gravitino_eigenvalues}
 a_{1\pm} &= \frac{1}{4\sqrt 3}\Sigma^2 \sqrt{2 \xi^{mn}\xi_{mn} \mp \lvert \tilde X \rvert } \nn \\
 & = \frac{1}{4\sqrt 3}\Sigma^2 \sqrt{2 \xi^{mn}\xi_{mn} \mp \sqrt{8(\xi^{mn}\xi_{mn})^2 -16\xi^{mn}\xi_{np}\xi^{pq}\xi_{qm}} }\, .
\end{align}
The index of the eigenvalues indicates the corresponding $SU(2)_\pm$ projected subspace. Recall that zero eigenvalues of
$\tensor{A}{_1_\, _i^j}$ are in one-to-one correspondence with conserved supersymmetries. From
\eqref{e:gravitino_eigenvalues} we see that there are three different possibilities for the amount of supersymmetry in
the vacuum:\footnote{Note that $\xi^{mn}\xi_{mn}$ is quadratic in each summand.}
\begin{center}
\begin{tabular}{|c|c|}
\hline
supersymmetry & condition \\
\hline\hline
\rule[-.2cm]{0mm}{.7cm} $\cN =4$ & $\xi^{mn} = 0 \quad \forall m,n$ \\
\hline
\rule[-.2cm]{0mm}{.7cm}  $\cN=2$ & $\xi^{mn}\xi_{np}\xi^{pq}\xi_{qm} = \frac{1}{4} (\xi^{mn}\xi_{mn})(\xi^{pq}\xi_{pq})$\\
\hline
\rule[-.2cm]{0mm}{.7cm}  $\cN=0$ & others \\
\hline
\end{tabular}
\end{center}

\section{Supergravity breaking from $\cN =4$ to $\cN = 2$} \label{N=4toN2breaking}

In this section we study the supersymmetry breaking from 
$\cN=4$ to $\cN=2$. 
We first comment further on the $\cN=2$ vacuum conditions 
in \autoref{N=2vacuum}.  This is followed by a discussion of the 
spectrum parameterizing fluctuations around this vacuum.  
More precisely, we focus on the fields that become massive 
by a Higgs mechanism. In \autoref{tensorialHiggs} we show
how some of the tensors become massive by eating 
vector degrees of freedom, a mechanism that we term 
\emph{tensorial Higgs mechanism}. The Higgs mechanism rendering 
half of the gravitinos massive is summarized in \autoref{superHiggs}.
Let us stress that both massive tensors and massive gravitinos 
will play a distinctive role in the evaluation of the one-loop effective
action presented in \autoref{sec:effectiveaction}. 

\subsection{Solution to the $\cN = 2$ condition} \label{N=2vacuum}

Since we are focusing on $\cN=2$ vacua in this work, we recall the corresponding 
supersymmetry condition from \autoref{SusyCond}. It takes the form 
\begin{align}
\label{e:half_susy_condition_1}
 \xi^{mn}\xi_{np}\xi^{pq}\xi_{qm} \overset{!}{=} \frac{1}{4} (\xi^{mn}\xi_{mn})(\xi^{pq}\xi_{pq}) 
 \quad \textrm{ and }\quad \exists\,  m,n \,\, \textrm{s.t.}\,\, \xi^{mn} \neq 0 \, .
\end{align}
where $\xi_{mn}$ is field dependent since it arises from the constant $\xi_{MN}$ via \eqref{e:dressed_gaugings}.
In this case the eigenvalues $a_{1\pm}$ of the gravitino mass matrix  given in \eqref{e:gravitino_eigenvalues}
take the form  
\begin{align} \label{N=2eigenvalues}
 a_{1+} = 0 \, , \qquad a_{1-} = \frac{1}{2\sqrt 3}\Sigma^2 \sqrt{\xi^{mn} \xi_{mn}} \, .
\end{align}
This implies that the unbroken R-symmetry is $SU(2)_+$ in \eqref{group-split}.

In the following we aim to find the general solution to \eqref{e:half_susy_condition_1}.
In order to do that we recall from \cite{Cassani:2012wc} that
$\xi^{mn}$ only acts on the orthogonal complement to $X^m$, i.e.~it satisfies the vanishing condition
\begin{align}
 \label{e:orth_property}
  X^m \xi_{mn} = 0 \, .
 \end{align}
We use the local $SO(5)$ symmetry in order to rotate
\begin{align}
 m \rightarrow (\tilde 0 , \hat m ) \, ,
\end{align}
where the index $\tilde 0$ refers to the direction of $X^m$.
Then \eqref{e:orth_property}
ensures that $\xi^{\tilde 0 m}=0$.
Note that we still retain the freedom of a reflection along the $X^m$ direction
as well as $SO(4)\subset SO(5)$ rotations orthogonal to $X^m$.
We continue
with the latter in order to bring the skew-symmetric
matrix $\xi^{\hat m \hat n}$ into block-diagonal form \cite{0521386322}
\begin{align} \label{splitximn}
 (\xi^{m  n})  
 = \begin{pmatrix}
 0 & 0 \\
 0 & \xi^{\hat m \hat n} \end{pmatrix}
 = \begin{pmatrix}
 0 & 0 & 0\\
 0 & \gamma \varepsilon &  0\\
 0 & 0 & \tilde \gamma \varepsilon
                         \end{pmatrix}\, ,
\end{align}
where $\gamma, \tilde \gamma \in \mathbb{R}$ and $\varepsilon$ denotes the usual two-dimensional epsilon tensor.
The values $\pm i \gamma \, , \pm i \tilde\gamma$ are the imaginary eigenvalues of $\xi^{mn}$.
Inserting this expression for $\xi^{mn}$ into \eqref{e:half_susy_condition_1} yields the condition
\begin{align}
 \gamma \overset{!}{=} \pm\tilde \gamma \, . 
\end{align}
Therefore the $\cN=2$ supersymmetry conditions can be summarized as
\begin{align}
 \cN = 2 \textrm{ SUSY}\quad \Leftrightarrow \quad (\xi^{mn}) \textrm{ has eigenvalues } 0\, , \pm i \gamma \, , \pm i \gamma \neq 0 \, .  
\end{align}

Note that we can employ the $SO(5)$ symmetry further to simplify the analysis. 
In fact, one can always choose
\begin{align}
 \gamma > 0 \qquad \textrm{and} \qquad \tilde \gamma = \gamma \, ,
\end{align}
which is possible since
\begin{align} 
\label{e:rotation1} \textrm{diag}(1,-1,1,-1,1) &\in SO(5) \, , \\
\label{e:rotation2} \textrm{diag}(-1,1,1,-1,1)  \, ,
 \textrm{diag}(-1,-1,1,1,1) &\in SO(5) \, 
\end{align}
are rotations that leave the $\xi^{\tilde 0 m}$ components invariant.
The transformation \eqref{e:rotation1} changes the absolute signs of $\gamma$ and $\tilde \gamma$,
while the transformations in \eqref{e:rotation2} change the relative sign between $\gamma$ and $\tilde \gamma$.
Note that \eqref{e:rotation1} leaves the direction of $X^m$ invariant, while the transformations \eqref{e:rotation2}
reflect along it.

\subsection{The tensorial Higgs mechanism} \label{tensorialHiggs}

In this subsection we introduce the tensorial Higgs mechanism that
is key to implementing the $\cN=4$ to $\cN=2$ supergravity breaking.
In order to do that we recall that the tensors appear in the combination \eqref{def-cH} as
\beq
  \cH^M_{\mu\nu}= F^M_{\mu\nu} - 2 \tensor{\xi}{_N^M}A_{[\mu}^0 A_{\nu]}^N + \frac{1}{2} \xi^{MN} B_{\mu\nu\, N}\, ,
\eeq
In order to perform the Higgs mechanism we apply the
gauge transformations \eqref{general_gaugetransform} for 
the perturbations around the $\cN=2$ vacuum. 
We will show explicitly which vector degrees of freedom get eaten by 
the tensors to render them massive.

Since the gauge transformation \eqref{general_gaugetransform} of $A^0$ is not dependent on the 
gauge parameters $ \Xi_{\mu\, N}$ of the tensors, $A^0$ will stay massless in the 
$\cN=2$ vacuum. In contrast, some of the $A^M_\mu$ can be absorbed by the 
tensors. To make this explicit, we introduce 
\begin{align} \label{rotate_AB}
&B_{\mu \nu\, m}  := \tensor{\langle \cV\rangle}{^M_m} B_{\mu \nu\, M} \, ,  
&& B_{\mu \nu\, a} := \tensor{\langle \cV\rangle}{^M_a} B_{\mu \nu \, M} \, , & \\
  &A_\mu^m := \tensor{\langle \cV\rangle}{_M^m} A_\mu^M \, ,  && A_\mu^a :=  \tensor{\langle\cV\rangle}{_M^a} A_\mu^M \, , & \nn
\end{align}
where the $\tensor{\langle \cV\rangle}{_M^m},\ \tensor{\langle \cV\rangle}{_M^a}$ are the vacuum expectation values (VEVs)
of the coset representatives  $\tensor{\cV}{_M^m}, \ \tensor{\cV}{_M^a}$ in the $\cN=2$ vacuum. Note that the 
position of the indices $M,N$ is crucial in these definitions.
Similarly, we can introduce the gauge parameters $(\Lambda^m,\Lambda^a)$ and
$(\Xi_{\mu \, n},\Xi_{\mu\, a} )$ by setting 
\begin{align}
   &\Lambda^m:=   \Lambda^M \tensor{\langle \cV\rangle}{_M^m} \,  , & & \Lambda^a  :=  \Lambda^M  \tensor{\langle  \cV\rangle}{_M^a} \, ,& \\
   &\Xi_{\mu \, m} :=   \Xi_{\mu \, M}  \tensor{\langle  \cV\rangle}{^M_m} \, , &&  \Xi_{\mu \, a} :=   \Xi_{\mu \, M}  
   \tensor{\langle  \cV\rangle}{^M_a}     \, .&    \nn
\end{align}
In this rotated basis
the gauge transformations  \eqref{general_gaugetransform} read 
\begin{align}
 \label{e:gauge_transf_1} 
 \delta A_\mu^m = & \partial_\mu \Lambda^m +  A_\mu^0 \tensor{\xi}{_p^m} \Lambda^p
 -\frac{1}{2}  \xi^{mn} \Xi_{\mu \, n} \\
 \label{e:gauge_transf_2} \delta A_\mu^a = & \partial_\mu \Lambda^a -  A_\mu^0 \tensor{\xi}{_b^a} \Lambda^b
 -\frac{1}{2}  \xi^{ab} \Xi_{\mu \, b}\, , 
\end{align}
where we have inserted the vacuum condition $\xi^{am}=0$.

In the next step we apply the $SO(5)\times SO(n)$ symmetry to rotate 
$\xi^{mn}$ and $\xi^{ab}$ into a convenient basis. For $\xi^{mn}$
we have done this already in \eqref{splitximn} and introduced the 
constant $\gamma$. Similarly, we now split 
\beq
     a \ \rightarrow \ (\tilde a,  \hat a)\ , \qquad \hat a = 1, \ldots, \text{rank}(\xi^{ab})\ .
\eeq
Since $\xi^{ab}$ is anti-symmetric one notes that  $\text{rank}(\xi^{ab})$ is 
even and we can define
\beq \label{def-nT}
    n_T := \frac{1}{2}  \text{rank}(\xi^{ab})\ . 
\eeq
Using the $SO(n)$ rotations one can then choose a basis such that $\xi^{ab}$ takes the 
form 
\begin{align} \label{splitxiab}
 (\xi^{a  b})  
 = \begin{pmatrix}
 0 & 0 \\
 0 & \xi^{\hat a \hat b} \end{pmatrix}
 = \begin{pmatrix}
 0 & 0 & 0 & 0\\
 0 & \gamma_1 \varepsilon &  \cdots & 0\\ 
 0 &  \vdots &  \ddots & \vdots\\
 0 & 0 & \cdots & \gamma_{n_T} \varepsilon
                         \end{pmatrix}\, ,
\end{align}
where $\varepsilon$ is the two-dimensional epsilon tensor. 
In this expression the non-zero real eigenvalues $\gamma_{\check a},\, \check a = 1,\ldots,n_T$ 
are parameterizing the non-trivial VEV of $\xi^{ab}$.

Together \eqref{splitximn} and \eqref{splitxiab} provide a diagonalization
of $\xi^{MN}$ after contraction with the VEVs $\langle \cV \rangle$. Recall, however, 
that the eigenvalues $\pm i \gamma,\pm i \gamma_{\check a}$ are identical to the 
eigenvalues of $\xi^{MN}$, since the contraction with $\langle \cV \rangle$
corresponds to a similarity transformation.

Using the  
explicit expressions \eqref{splitximn} and \eqref{splitxiab} in the gauge transformations 
\eqref{e:gauge_transf_1} and \eqref{e:gauge_transf_2} we first note that $\delta A_\mu^{\tilde 0}$ and $\delta A_\mu^{\tilde a}$
simply reduce
to 
\beq
   \delta A^{\tilde 0} = \partial_\mu \Lambda^{\tilde 0}\ ,\qquad \delta A_\mu^{\tilde a} = \partial_{\mu} \Lambda^{\tilde a}\ .
\eeq
These are the standard gauge transformations for $U(1)$ gauge fields. We conclude that 
the vector fields $(A^0,A^{\tilde 0},A^{\tilde a})$ remain massless and propagate in the 
$\cN=2$ effective theory describing the massless fluctuations around the vacuum.
Turning to the massive degrees of freedom, 
we can pick the gauge parameters $(\Xi_{\mu \, \hat n}, \Xi_{\mu \, \hat b})$ such that  
\begin{align} 
 \label{e:gauge_con} 
 A_\mu^{\hat m} \overset{!}{=} & \frac{1}{2}  \xi^{\hat m \hat n} \Xi_{\mu \, \hat n}\ , \qquad  
 A_\mu^{\hat a} \overset{!}{=}  \frac{1}{2}  \xi^{\hat a \hat b} \Xi_{\mu \, \hat b}\, .
\end{align}
By definition of $\xi^{\hat m \hat n}$ and $\xi^{\hat a \hat b}$ in \eqref{splitximn} and \eqref{splitxiab} both matrices have 
full rank and can be inverted to solve for $(\Xi_{\mu \, \hat n}, \Xi_{\mu \, \hat b})$. This implies that the 
vectors $A_\mu^{\hat m}$ and $A_\mu^{\hat a}$ are pure gauge and can be absorbed 
by the tensors $(B_{\mu \nu \, \hat m},B_{\mu \nu \, \hat a})$ to render them massive.

In computing the $\cN=2$ effective action for the massless degrees
of freedom it will be crucial to include one-loop effects of massive 
tensors. Therefore it is necessary to evaluate the action 
of the massive tensors $(B_{\mu \nu \, \hat m},B_{\mu \nu \, \hat a})$ after the gauge-fixing \eqref{e:gauge_con}.
Inserting the definitions \eqref{e:dressed_gaugings}, \eqref{rotate_AB}, and the split form of $\xi^{mn}$ and $\xi^{ab}$ given in 
the first equalities of \eqref{splitximn}, \eqref{splitxiab}  
into the general $\cN=4$ action \eqref{bos_N=4action} we extract all terms depending on the tensors.
Explicitly, after gauging away the vectors $A_\mu^{\hat m},A_\mu^{\hat a}$, they read
\begin{align}
 \label{e:tensor_act_1}
 e^{-1} \cL_{B}  =&  \frac{1}{16 \sqrt 2}\epsilon^{\mu\nu\rho\lambda\sigma}  \Big \lbrace
  \xi^{\hat m \hat n} B_{\mu\nu\, \hat m} \Big [ \partial_\rho B_{\lambda\sigma\, \hat n}
 -  \tensor{\xi}{^{\hat p}_{\hat n}}A^0_\rho B_{\lambda\sigma\, \hat p} \Big ]  
 \\ &\qquad \qquad \qquad 
 +\xi^{\hat a \hat b} B_{\mu\nu\, \hat a} \Big [ \partial_\rho B_{\lambda\sigma\, \hat b}
 -  \tensor{\xi}{^{\hat c}_{\hat b}}A^0_\rho B_{\lambda\sigma\, \hat c} \Big ] \Big \rbrace \nn   \\
 &- \frac{1}{16} \Sigma^2  \xi^{\hat m \hat p} \xi_{\hat m \hat q} B_{\mu\nu \, \hat p} B^{\mu\nu \, \hat q}
 - \frac{1}{16} \Sigma^2  \xi^{\hat a \hat b} \xi_{\hat a \hat c} B_{\mu\nu \, \hat b} B^{\mu\nu \, \hat c} \, .\nn
\end{align}
The first two lines in this expression are the kinetic terms of $(B_{\mu \nu \, \hat m},B_{\mu \nu \, \hat a})$,
while the last line summarizes their mass terms. Since $\xi^{\hat m \hat n}$ and $\xi^{\hat a\hat b}$
have maximal rank, indeed all such tensors are massive. 

The Lagrangian \eqref{e:tensor_act_1} can be further simplified by going to the basis in which 
$ \xi^{\hat m \hat n},  \xi^{\hat a \hat b}$ are parametrized by 
the eigenvalues $\gamma,\gamma_{\check a}$ as in \eqref{splitximn}, \eqref{splitxiab}. 
The appearance of the two-dimensional epsilon tensor in these expressions 
makes it natural to define the \emph{complex} tensors
\begin{align} \label{complex-tensors}
 &\BB_{\alpha} := B_{2\alpha-1} + i B_{2\alpha}\ , &\alpha = 1,2  && (\text{use}  \ \, B_{\hat m})& \\
 &\BB_{\check a} := B_{2\check a-1} + i B_{2\check a } \, , & \check a = 1,\ldots, n_T& &(\text{use} \ \, B_{\hat a}) & \nn 
\end{align}
with $n_T$ defined in \eqref{def-nT}. 
One can show that the $\alpha$ index corresponds to the fundamental representation of the $\cN=2$ R-symmetry group $SU(2)$. 
Here and in the following we will use boldface symbols to denote complex fields.
Inserting these definitions 
together with  \eqref{splitximn} and \eqref{splitxiab} into \eqref{e:tensor_act_1}
we find\footnote{In analogy
to the fermions we define $\bar\BB^\alpha := (\BB_\alpha)^*\,$.}
\begin{align}
\label{e:tensor_act_2}
 e^{-1} \cL_{B} = 
 &- \frac{1}{16} \Big[  i  \frac{1}{ \sqrt 2}\epsilon^{\mu\nu\rho\lambda\sigma}  \gamma 
 \bar \BB_{\mu\nu}^\alpha  ( \partial_\rho   \tensor{\BB}{_\lambda_\sigma_\, _{\alpha}} 
 - i  \gamma \tensor{\BB}{_\lambda_\sigma_\,_{\alpha}} A^0_\rho ) + \Sigma^2 \gamma^2 \bar \BB_{\mu\nu}^\alpha \BB^{\mu\nu}_{\alpha} \Big] \nn \\
 &-\frac{1}{16}  \sum_{\check a} \Big[  i   \frac{1}{\sqrt 2}\epsilon^{\mu\nu\rho\lambda\sigma} \gamma_{\check a} 
 \tensor{\bar \BB}{_\mu_\nu_\, _{\check a}}  ( \partial_\rho   \tensor{\BB}{_\lambda_\sigma_\, _{\check a}} 
 - i  \gamma_{\check a} \tensor{\BB}{_\lambda_\sigma_\,_{\check a}} A^0_\rho )  
 + \Sigma^2   \gamma_{\check a}^2 \bar \BB_{\mu\nu\, \check a} \BB^{\mu\nu}_{\check a} \Big]\, .
\end{align}

In the last step we want to rescale the complex tensors in order to 
bring the action into the standard form 
\begin{align} \label{standard_complex_tensor}
  e^{-1}\cL_B &= -\frac{1}{4}i c_\BB\, \epsilon^{\mu\nu\rho\sigma\tau}\bar \BB_{\mu\nu}\cD_\rho \BB_{\sigma\tau} 
 -\frac{1}{2}m_\BB\, \bar \BB_{\mu\nu} \BB^{\mu\nu}\, , 
\end{align}
with $\cD_\rho \BB_{\sigma\tau} =   \partial_\rho \BB_{\sigma\tau} - i q_\BB\, A^0_\rho\, \BB_{\sigma\tau}$. 
This Lagrangian was used in the one-loop computations of \cite{Bonetti:2012fn,Bonetti:2013ela}. Here $m_\BB>0$ is the real mass of the complex 
tensor $\BB_{\mu \nu}$ and $q_\BB$ encodes its charge under the $U(1)$ vector $A^0$.
The choice of $c_\BB $ is dependent on the representation of $\BB_{\mu \nu}$
under the massive little group $SO(4)$ in five dimensions.
Explicitly, one has
\begin{align} \label{cBB_choice}
 c_{\BB} &= + 1 \Leftrightarrow (1,0 )\textrm{ of } SO(4) \\
 c_{\BB} &= - 1 \Leftrightarrow ( 0,1 ) \textrm{ of } SO(4)  \, . \nn
\end{align}
Comparing \eqref{e:tensor_act_2} with \eqref{standard_complex_tensor} we
can determine $c_\BB,m_\BB,q_\BB$ after rescaling 
\begin{align}
 \big( \BB_\alpha  , \BB_{\check a} \big )
 \mapsto \frac{1}{2^{5/4}}
 \bigg(\sqrt \gamma  \BB_\alpha  , \sqrt{\vert\gamma_{\check a}\vert} \BB_{\check a}\bigg )\, . 
\end{align}
This results in the identifications
\begin{subequations}\label{tensor_data}
\begin{align} 
 &c_{\BB_\alpha} = 1 \, , && m_{\BB_\alpha} =\frac{1}{\sqrt 2}\Sigma^2 \gamma \, ,&& q_{\BB_{ \alpha}} =  \gamma \, , \\
 & c_{\BB_{\check a}}= \sign \gamma_{\check a}\, ,&&m_{\BB_{\check a}}=  \frac{1}{\sqrt 2}\Sigma^2  \vert \gamma_{\check a} \vert \, ,
 && q_{\BB_{\check a}} =  \gamma_{\check a}\, . &
\end{align}
\end{subequations}
This concludes our discussion of the massive tensors. We found that 
evaluated around the $\cN=2$ vacuum there are $n_T+2$ complex 
massive tensors $(\BB_\alpha,\BB_{\check a})$ with 
standard action \eqref{standard_complex_tensor} and characteristic 
data \eqref{tensor_data}.
For convenience we summarize the 
split of the fields induced by $\xi^{MN}$ in \autoref{index_split}.
\begin{table}[h!]
 \centering
\begin{tabular}{|c|c|c|c|c|}
\hline
  \rule[-.2cm]{0cm}{.7cm} & rotation with $\langle \cV \rangle$ &  $\xi^{MN}$-split & physical degrees\\
\hline\hline
\multirow{4}{*}{$(A^M,B_M)$} & \multirow{2}{*}{$(A^m,B_m)$} & \rule[-.2cm]{0cm}{.7cm}$(A^{\tilde 0},B_{\tilde 0})$ &  $A^{\tilde 0}$  massless \\ 
   \cline{3-4}
   & & \rule[-.2cm]{0cm}{.7cm}$(A^{\hat m},B_{\hat m})$ &  $\BB_{\alpha}$ complex, massive\\
   \cline{2-4}
   & \multirow{2}{*}{$(A^a,B_a)$} & \rule[-.2cm]{0cm}{.7cm}$(A^{\tilde a},B_{\tilde a})$  & $A^{\tilde a}$ massless\\  
   \cline{3-4}
    & & \rule[-.2cm]{0cm}{.7cm}$(A^{\hat a},B_{\hat a})$ & $\BB_{\check a}$ complex, massive\\
\hline
\end{tabular}
\caption{Natural split of $A^M$ and $B_M$ induced by $\xi^{MN}$.}
\label{index_split}
\end{table}

\subsection{The super-Higgs mechanism} \label{superHiggs}

In the $\cN =2$ broken phase of a $\cN=4$ theory a 
gravitino mass term has to be generated for half of the 
gravitino degrees of freedom. 
This mass arises in the sector of the broken $SU(2)$ subgroup of the $\cN=4$ R-symmetry group $USp(4)$. In fact, 
two gravitini eat up two spin-1/2 goldstini from the gravity multiplet and become massive.
In this super-Higgs mechanism the massive gravitini acquire four extra degrees of freedom.
The appropriate description of the massive fields is in
terms of a single Dirac spin-3/2 fermion $\Bpsi_\mu$ without a symplectic Majorana condition.
The massive gravitino combines with the two massive complex tensors $\BB_\alpha$ of the gravity multiplet into
a massive $\cN = 2$ gravitino multiplet $(\Bpsi_\mu,\BB_\alpha)$. The construction of such a half-BPS multiplet 
has been discussed in \cite{Hull:2000cf}.
In the following 
we will briefly discuss the super-Higgs mechanism and determine the mass and $U(1)$ charge of the 
gravitino multiplet. 

Let us first consider the four $\cN =4$ symplectic Majorana gravitini $\psi_\mu^i$
and the spin-1/2 fermions in the gravity multiplet $\chi^i$. These split under the breaking
\begin{align}
 USp(4) \rightarrow SU(2)_+ \times SU(2)_- \, .
\end{align}
into $\psi_\mu^\alpha$, $\psi_\mu^{\dot\alpha}$ and $\chi^\alpha$, $\chi^{\dot\alpha}$, respectively. As noted in the
last subsection, the index $\alpha = 1,2$ refers to the fundamental representation of the $\cN = 2$ R-symmetry
group $SU(2)_+$, while $\dot \alpha = 1,2$ corresponds to the broken $SU(2)_-$ part. Both indices are raised and lowered
with the epsilon tensor analogous to \eqref{e:properties_omega} and \eqref{e:raising_lowering}.
As one can see in \eqref{fermionic-quadratic} using the $\cN = 2$ vacuum conditions \eqref{e:vacuum_condition} and \eqref{e:half_susy_condition_1}
all fermion bilinears involving
$\psi_\mu^\alpha$ and $\chi^\alpha$ vanish in the vacuum,
leaving only the kinetic terms for these fields.
Thus we find two massless $\cN = 2$ spin-3/2 symplectic Majorana fermions $\psi_\mu^\alpha$ and two
massless spin-1/2 symplectic Majorana fermions $\chi^\alpha$.
The reduced symplectic Majorana condition reads
\begin{align}
 \bar \chi^\alpha := ( \chi_\alpha )^\dagger \gamma_0 = \epsilon^{\alpha\beta} \chi_\beta^T C \,
\end{align}
and similarly for $\psi_\mu^\alpha$, where
$C$ denotes the charge conjugation matrix and $\epsilon^{\alpha\beta}$ is the 
two-dimensional Levi-Civita symbol with $\epsilon^{12}=+1$.
We note that throughout this paper
all massless fermionic $\cN = 2$ fields are taken to be symplectic Majorana.

We proceed with the investigation of the remaining fields $\psi_\mu^{\dot\alpha}$ and
$\chi^{\dot\alpha}$. The terms in the Lagrangian involving these fields are
(ignoring fluctuations of scalars)
\begin{align}
 e^{-1}\cL_{\psi , \chi}=&
 -\frac{1}{2} \bar \psi^{\dot\alpha}_\mu \gamma^{\mu\nu\rho}\cD_\nu \psi_{\rho\, \dot\alpha}
 -\frac{1}{2}\bar\chi^{\dot\alpha} \slashed\cD \chi_{\dot\alpha}
 +\frac{\sqrt 6 i }{4} \langle A_1 \rangle_{\dot\alpha\dot\beta}
 \bar \psi^{\dot\alpha}_\mu \gamma^{\mu\nu} 
 \psi^{\dot\beta}_\nu  \nn \\
& - \frac{5\sqrt 6 i}{12}  \langle A_1 \rangle_{\dot\alpha\dot\beta}
\bar\chi^{\dot\alpha} \chi^{\dot\beta}
 - \sqrt 2  \langle A_1 \rangle_{\dot\alpha\dot\beta}
 \bar\psi^{\dot\alpha}_\mu \gamma^\mu \chi^{\dot\beta}\, \nn\\
 = &  -\frac{1}{2} \bar \psi^{\dot\alpha}_\mu \gamma^{\mu\nu\rho}\cD_\nu \psi_{\rho\, \dot\alpha}
 -\frac{1}{2}\bar\chi^{\dot\alpha} \slashed\cD \chi_{\dot\alpha}
 -\frac{1}{2\sqrt 2}\Sigma^2  \gamma
 \bar \psi^{\dot\alpha}_\mu \gamma^{\mu\nu} 
 \psi_{\nu\, \dot \alpha}  \nn \\
& - \frac{5\sqrt 2}{12}\Sigma^2  \gamma  \bar\chi^{\dot\alpha} \chi_{\dot\alpha}
 - \frac{\sqrt 6}{3} \Sigma^2 \gamma
 \bar\psi^{\dot\alpha}_\mu \gamma^\mu \chi_{\dot\alpha} \, ,
\end{align}
where we used that due to \eqref{N=2eigenvalues} the two eigenvalues $\pm a_{1-}$ of 
$\tensor{\langle A_1 \rangle}{_{\dot\alpha}^{\dot\beta}}$ are given by
\begin{align}
 a_{1-} = \frac{1}{\sqrt 3}\Sigma^2 \gamma \, .
\end{align}
The $\chi^{\dot\alpha}$ actually are the goldstini, that render the
$\psi^{\dot\alpha}_\mu$ massive, and can be removed from the
action by a shift of the gravitini analog to the one performed in \cite{Hohm:2004rc,Horst:2012ub}.
We thus obtain
\begin{align}
 e^{-1}\cL_{\textrm{mass grav}}=&
 -\frac{1}{2} \bar \psi^{\dot\alpha}_\mu \gamma^{\mu\nu\rho}\cD_\nu \psi_{\dot\alpha\, \rho}
 - \frac{1}{2}\frac{1}{\sqrt 2}\Sigma^2 \, \gamma\,
 \bar \psi^{\dot\alpha}_\mu \gamma^{\mu\nu} 
 \psi_{\dot \alpha \, \nu}\, .
\end{align}
It is now convenient to merge the two symplectic Majorana fermions into a single
unconstrained Dirac spinor
\footnote{We could also choose $\Bpsi_\mu := \psi_\mu^{\dot\alpha = 2}$,
which flips the representation and the charge under $A^0_\mu$, since both descriptions
are equivalent.}
\begin{align} \label{def-Bpsi}
 \Bpsi_\mu := \psi_\mu^{\dot\alpha = 1} \, .
\end{align}
The action then reads
\begin{align} \label{mass_gravitino}
 e^{-1}\cL_{\textrm{mass grav}}=&
 -\bar \Bpsi_\mu \gamma^{\mu\nu\rho}\cD_\nu \Bpsi_{\rho}
 - \frac{1}{\sqrt 2}\Sigma^2 \, \gamma \,
 \bar \Bpsi_\mu \gamma^{\mu\nu} 
 \Bpsi_{\nu}\, ,
\end{align}
with $\cD_\mu \Bpsi_\nu = \partial_\mu \Bpsi_\nu 
 - i  \gamma A^0_\mu  \Bpsi_\nu$.

To conclude this section we compare the action \eqref{mass_gravitino} with 
the standard form 
\begin{align}
 \label{e:lagrangian_gravitino}
 e^{-1} \cL_\Bpsi = - \bar \Bpsi_\mu \gamma^{\mu\nu\rho}  \cD_\nu \Bpsi_{\rho} - 
  c_{\Bpsi}\,  m_\Bpsi \, 
 \bar \Bpsi_\mu \gamma^{\mu\nu} \psi_{\nu }
 \, , 
\end{align}
where $\cD_\nu \Bpsi_{\rho} = \partial_\nu \Bpsi_{\rho} - i q_\Bpsi A_{\nu} \Bpsi_\rho$ and  
$c_{\Bpsi}$ depends on the representation under the massive little group $SO(4)$ as
\begin{align} \label{cBpsi_choice}
 c_{\Bpsi} &= + 1 \Leftrightarrow (1 , \tfrac{1}{2})\textrm{ of } SO(4)\, ,  \\
 c_{\Bpsi} &= - 1 \Leftrightarrow (\tfrac{1}{2}, 1 )\textrm{ of } SO(4)  \, . \nn
\end{align}
Comparing \eqref{mass_gravitino} with \eqref{e:lagrangian_gravitino} we 
conclude that $\Bpsi$ is in the $(1, \frac{1}{2})$ representation of
$SO(4)$ and carries mass and $A^0_\mu$-charge
\begin{align}
 c_{\Bpsi} =1 \, ,\qquad m_\Bpsi = \frac{1}{\sqrt 2}\Sigma^2  \gamma \, , \qquad 
 q_\Bpsi  =  \gamma \, .
\end{align}
These data will be crucial in evaluating the one-loop corrections induced 
by the massive gravitino multiplet in the next section. 

The massive Dirac gravitino $\Bpsi$ combines with the massive tensors $\BB_\alpha$
of the last section into a massive gravitino multiplet.

\section{Mass spectrum and effective action} \label{sec:effectiveaction}

In this section we identify the complete spectrum parameterizing the fluctuations around 
the $\cN=2$  vacuum. We determine the masses and $U(1)$ charges of all fields
and show how they reassemble in $\cN = 2$ multiplets in \autoref{N=2spectrum}.
Furthermore, we derive the low-energy effective action of the massless modes 
with particular focus on the data determining the $\cN=2$ vector sector. 
The classical truncation from $\cN=4$ to $\cN=2$ is discussed in \autoref{Classicaltruncation}. 
The crucial inclusion of one-loop quantum corrections due to integrating out 
massive fermions and tensors is discussed in \autoref{oneloopeffects}. 
These induce extra contributions to the metric and Chern-Simons terms that are independent of 
the scale of supersymmetry breaking.

\subsection{The $\cN = 2$ spectrum} \label{N=2spectrum}

The $\cN = 2$ spectrum and its properties can be determined by evaluating 
the $\cN=4$ action in the vicinity of the $\cN=2$ vacuum. 
To read off the masses and charges all kinetic terms and mass terms have 
to be brought into canonical form after spontaneous symmetry breaking. 
This diagonalization procedure is rather lengthy and therefore partially deferred
to \autoref{mass_appendix} in detail. In the following we highlight 
some of the basic steps and summarize the results.

The key ingredients in the mass generation are the gaugings $\xi^{MN}$.
Recall that in the background we rotated $\xi^{MN}$ to $\xi^{mn},\xi^{ab}$
and found the components
\begin{align}
   \xi^{mn} & \rightarrow  \xi^{\hat m \hat n}\, ,& \xi^{\tilde 0 \hat n}& =  \xi^{\tilde 0 \tilde 0} = 0 \, ,\\ 
   \xi^{ab} &  \rightarrow  \xi^{\hat a \hat b}\, ,& \xi^{\tilde a \hat b}& =  \xi^{\tilde a \tilde b} = 0 \, ,
\end{align}
where $\xi^{\hat m \hat n}$ and $ \xi^{\hat a \hat b}$ have maximal rank.
This yielded the natural index split
\begin{align}\label{e:split}
\begin{array}{ccccc}
 m &\rightarrow & (\tilde 0 , \hat m) &\rightarrow & (\tilde 0 , [\alpha 1] , [\alpha 2])\, ,\\ 
 a &\rightarrow & (\tilde a , \hat a ) &\rightarrow & (\tilde a , [\check a 1],[\check a 2] )\, . 
\end{array}
\end{align}
Here the splits of $\hat m $ into $[\alpha 1],[\alpha 2]$ arises due to the block-diagonalization in \eqref{splitximn}
with the first index $\alpha$ labeling the two blocks and the second index labeling the two entries of each block. The split of $ \hat a$ into $[\check a 1],[\check a 2]$ arises 
due to the blocks in \eqref{splitxiab}. 

In order to extract the massless and massive scalar spectrum
it is convenient to reparameterize the element $\cV$ of the coset space $\cM_{5,n}$ introduced in \eqref{coset_def}.
Denoting  by $\langle \cV \rangle$ the value of $\cV$ in the $\cN=2$ vacuum as in \eqref{rotate_AB}, 
we write 
\beq \label{cVexpansion}
     \cV  = \langle \cV \rangle\, \text{exp} \big( \phi^{ma} [t_{ma}]\big)\ ,
\eeq
where $[t_{ma}]_M^{\ N} = 2 \delta_{[m}^{\ \ N} \eta_{a]M}$. The $\phi^{ma}$ then correspond to 
the unconstrained fluctuations around the vacuum value 
$ \langle \cV \rangle$ and constitute the scalar degrees of freedom in the $\cN = 2$ effective theory. 
Due to the index split \eqref{e:split} we need to consider the scalars 
\beq \label{splittingphi}
  \phi^{ma}\ \rightarrow \ (\phi^{\tilde 0 \tilde a}, \phi^{\tilde 0 [\check a 1]},\phi^{\tilde 0 [\check a 2]}) \ (\phi^{[\alpha 1] \tilde a},\phi^{[\alpha 1] [\check a 1]},\phi^{[\alpha 1] [\check a 2]} )\ 
  (\phi^{[\alpha 2] \tilde a}, \phi^{[\alpha 2] [\check a 1]},\phi^{[\alpha 2] [\check a 2]})\ .
\eeq
To treat these more compactly we introduce, just as for tensors in \eqref{complex-tensors}, the complex scalars
\begin{subequations}\label{compl_scalars}
\bea 
    \Bphi^{\alpha \tilde a} &:=& \tfrac{1}{\sqrt{2}} (\phi^{[\alpha1] \tilde a} + i \phi^{[\alpha 2] \tilde a} )  \ ,\qquad 
    \Bphi^{\tilde 0 \hat a} := \tfrac{1}{\sqrt{2}} (\phi^{\tilde 0 [\check a 1]}+i\phi^{\tilde 0 [\check a 2]})\, , \label{compl_scalars_1} \\
   \Bphi^{\alpha \check a}_1 &:=& \tfrac{1}{2} (\phi^{[\alpha 1] [\check a 1]} - \phi^{[\alpha 2] [\check a 2]}
   + i \phi^{[\alpha 2] [\check a 1]} + i \phi^{[\alpha 1] [\check a 2]}) \label{compl_scalars_2}\\
   \Bphi^{\alpha \check a}_2 &:=& \tfrac{1}{2} (\phi^{[\alpha 1] [\check a 2]} - \phi^{[\alpha 2] [\check a 1]}
   + i \phi^{[\alpha 2] [\check a 2]} + i \phi^{[\alpha 1] [\check a 1]}) \, .\label{compl_scalars_3}
\eea
\end{subequations}
Note that in this way all $\phi^{ma}$ of  the split \eqref{splittingphi} except $\phi^{\tilde 0 \tilde a}$
are combined into complex scalars. 

Similarly, we proceed for the split of the $\cN=4$ fermions $\lambda^{a}_i$. Note 
that as for the gravitino in \autoref{superHiggs} one splits $i \rightarrow (\alpha, \dot \alpha)$.
Together with the index split of $a$ given in \eqref{e:split} one has 
\beq \label{split_lambdas}
   \lambda^{a}_i \ \rightarrow \ (\lambda^{\tilde a}_\alpha,\lambda_\alpha^{[\check a 1]},\lambda_\alpha^{[\check a 2]}) \ (\lambda_{\dot \alpha}^{ \tilde a},\lambda_{\dot \alpha}^{[\check a 1]},\lambda_{\dot \alpha}^{[\check a 2]})\ .
\eeq
It turns out to be convenient to combine all $\lambda^a_i$ except $\lambda^{\tilde a}_\alpha$ 
into complex Dirac fermions
\begin{subequations}\label{compl_fermions}
\bea \label{compl_fermions_1}
   \Blambda^{\check a}_\alpha &:=& \tfrac{1}{\sqrt{2}} (\lambda^{[\check a1]}_\alpha+ i \lambda^{[\check a2]}_\alpha)\ ,\\
   \label{compl_fermions_2}
  \Blambda^{\tilde a} &:=& \lambda^{\tilde a}_{\dot \alpha =1} \ , \quad \Blambda^{\check a}_1 := \frac{1}{\sqrt 2}(\lambda_{\dot \alpha =1}^{[\check a 1]} + i \lambda_{\dot \alpha =1}^{[\check a 2]})\, , \qquad 
 \Blambda^{\check a}_2 := \frac{1}{\sqrt 2}(\lambda_{\dot \alpha =1}^{[\check a 1]} - i \lambda^{[\check a 2]}_{\dot \alpha=1})\, . 
\eea
\end{subequations}
To justify the use of \eqref{compl_fermions} we stress that the appearance of all spin-1/2 fermions
can be expressed in terms of the unconstrained Dirac spinors $\Blambda^{\check a}_\alpha$, $\Blambda^{\tilde a}$, 
$\Blambda^{\check a}_{1,2}$.
Concerning \eqref{compl_fermions_1}, the other linear combination
$\tfrac{1}{\sqrt{2}} (\lambda^{[\check a1]}_\alpha - i \lambda^{[\check a2]}_\alpha)$
is related to $\Blambda^{\check a}_\alpha$ by the symplectic Majorana condition.
In \eqref{compl_fermions_2}, by the same reasoning, the linear combinations with $\lambda^a_{\dot \alpha = 2}$ are related to those
involving $\lambda^a_{\dot \alpha = 1}$.
All degrees of freedom of the massive spin-1/2 fermions are therefore captured
by the spinors \eqref{compl_fermions}, dropping the symplectic Majorana condition.

We are now in the position to summarize the spectrum. 
From the $\cN=4$ gravity multiplet the metric $g_{\mu \nu}$, two 
gravitini $\psi^\alpha_\mu$, two spin-1/2 fermions $\chi_\alpha$, two
vectors $A^0,A^{\tilde 0}$, and one scalar $\Sigma$ remain massless.
These fields group into the $\cN=2$ gravity multiplet $(g_{\mu \nu}, A^{\tilde 0},\psi^\alpha_\mu)$
and one $\cN=2$ vector multiplet $(A^0,\Sigma,\chi_\alpha)$. Note that the vector multiplet $(A^0,\Sigma,\chi_\alpha)$ is
special, since the massive states, such as the tensors and gravitini discussed in \autoref{tensorialHiggs} and \autoref{superHiggs}, carry $A^0$ charge.
In order to later derive the quantum effective action for the $A^0$ vector multiplet 
we need to determine the $A^0$-charge of all massive states. 
Additional We stress that the identifications in \eqref{compl_fermions_2} are analogous to the definition of the massive gravitino \eqref{def-Bpsi}.
$n-2n_T$ vector multiplets $(A^{\tilde a},\phi^{\tilde 0 \tilde a},\lambda_\alpha^{\tilde a})$ remain massless. 
We have already discussed the massless vectors $A^{\tilde a}$ in \autoref{tensorialHiggs}. 
Inserting \eqref{cVexpansion} into the $\cN=4$ action we 
check in \autoref{mass_appendix} that the $ \phi^{\tilde 0 \tilde a}$ and $\lambda_\alpha^{\tilde a}$ are indeed 
massless scalars. 

Recall that an $\cN=2$ hypermultiplet has four real scalars and one Dirac spin-1/2 fermion. Using the 
above definitions \eqref{compl_scalars} and \eqref{compl_fermions} one can form the hypermultiplets
\beq 
   (\Bphi^{\alpha \tilde a},\Blambda^{\tilde a})\, , \quad (\Bphi^{\alpha \check a}_1, \Blambda^{\check a}_1)\, , \quad  (\Bphi^{\alpha \check a}_2, \Blambda^{\check a}_2)
\eeq
The $n-2n_T$ hypermultiplets $(\Bphi^{\alpha \tilde a},\Blambda^{\tilde a})$ are always massive, since they receive masses 
$m_{\tilde a} = \frac{1}{\sqrt 2} \Sigma^2 \gamma$
from a non-trivial $\xi^{\hat m \hat n}$. The hypermultiplets $ (\Bphi^{\alpha \check a}_{1,2}, \Blambda^{\check a}_{1,2})$
can be either massless or massive, since their masses have two contributions
from a non-trivial $\xi^{\hat m \hat n}$ and $\xi^{\hat a \hat b}$, respectively.
As we show in \autoref{mass_appendix} the $\xi^{MN}$-splits \eqref{splitximn} and \eqref{splitxiab} yield
masses given by
\begin{align} \label{mass_general}
 m_{\check a}^1 = \frac{1}{\sqrt 2}\Sigma^2  \vert  \gamma  - \gamma_{\check a}\vert  \, ,\qquad m_{\check a}^2 = \frac{1}{\sqrt 2}\Sigma^2  \vert \gamma + \gamma_{\check a}  \vert \, ,
\end{align}
for the fields $(\Bphi^{\alpha \check a}_{1}, \Blambda^{\check a}_{1})$ and $(\Bphi^{\alpha \check a}_{2}, \Blambda^{\check a}_{2})$, respectively.
This implies that one hypermultiplets is massless whenever 
the condition
\beq  \label{massless_condition}
   \gamma_{\check a} = \gamma \qquad \text{or} \qquad \gamma_{\check a} = - \gamma 
\eeq
is satisfied. We 
denote the number of such massless hypermultiplets by $n_H$, and name them
$(h^\Lambda_{1,2,3,4}, \lambda^\Lambda_\alpha)$, with $\Lambda = 1,\ldots, n_H$. 
Due to the fact that the hypermultiplets appear in pairs, the existence 
of a massless hypermultiplet implies the existence of a massive hypermultiplet with 
mass $\sqrt 2\Sigma^2  \gamma$. Furthermore, one can check that 
one can consistently choose all $\gamma_{\check a}>0$ without 
changing the effective theory.
In summary, one has  $2n_T - n_H$ massive hypermultiplets with mass \eqref{mass_general}
out of the set $(\Bphi^{\alpha \check a}_{1,2}, \Blambda^{\check a}_{1,2})$.
Together with the $(\Bphi^{\alpha \tilde a},\Blambda^{\tilde a})$ one finds a total of $n-n_H$ massive hypermultiplets.

To complete the summary of the spectrum 
recall that in \autoref{tensorialHiggs} and \autoref{superHiggs} we have already identified and analyzed the
$\cN = 2$ massive gravitino multiplet comprising a massive Dirac gravitino $\Bpsi_\mu$ and two
complex massive tensors $\BB_\alpha$. 
Furthermore, we found $n_T$ complex massive tensors $\BB_{\check a}$. 
The latter combine with Dirac fermions $\Blambda^{\check a}_\alpha$ into $n_T$ complex massive tensor multiplets.

To conclude we list in \autoref{field_decomposition} the decompositions 
of the $\cN = 4$ fields in terms of $\cN =2$ fields along with their masses and charges. 
The reorganization into $\cN=2$ multiplets can be found in \autoref{vacuum_multiplets}.

\setlength\extrarowheight{5pt}
\begin{table}[]
\centering
\begin{tabular}{|c||c|l|l|l|}
\hline
$\cN = 4$  & $\cN = 2$ & mass & $c_{\rm field}$, SO(4) rep& charge \\
fields & fields & &for massive fields& under $A^0$\\
\hline\hline
\rule[-.3cm]{0cm}{.8cm}  $g_{\mu\nu}$ & $g_{\mu\nu}$ &0& - & 0\\
\hline
\rule[-.3cm]{0cm}{.8cm}   $A^0_\mu$ & $A^0_\mu$ &0& - & 0\\
\hline
\rule[-.3cm]{0cm}{.8cm}   $A_\mu^{ij}$ & $A^{\tilde 0}_\mu$ &0 &- & 0\\
 \rule[-.3cm]{0cm}{.8cm}  & $\BB_{\mu\nu\,\alpha}$ & $\frac{1}{\sqrt 2} \Sigma^2  \gamma$ & 1 & $ \gamma$ \\
\hline
 \rule[-.3cm]{0cm}{.8cm}  $\psi_\mu^i$ & $\psi_\mu^\alpha$ & 0 &-& 0\\
 \rule[-.3cm]{0cm}{.8cm} & $\Bpsi_\mu$ & $\frac{1}{\sqrt 2} \Sigma^2  \gamma$ & 1 & $ \gamma$ \\
\hline
 \rule[-.3cm]{0cm}{.8cm} $\chi_i$ & $\chi_\alpha$ & 0 & -& 0\\
 \rule[-.3cm]{0cm}{.8cm} & $\chi_{\dot\alpha}$ &-& goldstino & -\\
\hline
\rule[-.3cm]{0cm}{.8cm}  $\Sigma$ & $\Sigma$ & 0 &-& 0\\
\hline
\rule[-.3cm]{0cm}{.8cm}  $A_\mu^a$ & $A_\mu^{\tilde a}$ &0 & -& 0\\
\rule[-.3cm]{0cm}{.8cm} & $\BB_{\mu\nu\, \check a}$ & $\frac{1}{\sqrt 2}\Sigma^2  \vert \gamma_{\check a} \vert$
& $\text{sign}(\gamma_{\check a})$  & $\gamma_{\check a}$ \\
\hline
\rule[-.3cm]{0cm}{.8cm} $\lambda^a_i$ & $\lambda^{\tilde a}_\alpha$ & 0 & -& 0\\
\rule[-.3cm]{0cm}{.8cm}  & $\Blambda^{\check a}_\alpha$ & $\frac{1}{\sqrt 2}\Sigma^2  \vert\gamma_{\check a}\vert$ 
& $\text{sign}(\gamma_{\check a})$& $\gamma_{\check a}$ \\
\rule[-.3cm]{0cm}{.8cm}  & $\Blambda^{\tilde a}$ & $\frac{1}{\sqrt 2} \Sigma^2  \gamma$ & -1& $ \gamma$ \\
\rule[-.3cm]{0cm}{.8cm}  & $\Blambda^{\check a}_{1,2}$ & $\frac{1}{\sqrt 2}\Sigma^2  \vert \gamma \mp \gamma_{\check a}  \vert$ 
& $\text{sign}(\gamma \mp \gamma_{\check a} )$  
& $ \gamma \mp \gamma_{\check a}   $\\
\hline
\rule[-.3cm]{0cm}{.8cm}  $\phi^{m a}$ & $\phi^{\tilde 0 \tilde a}$ & 0  & -& 0\\
\rule[-.3cm]{0cm}{.8cm}  & $\Bphi^{\alpha\tilde a}$ & $\frac{1}{\sqrt 2} \Sigma^2  \gamma$ & singlet & $\gamma$ \\
\rule[-.3cm]{0cm}{.8cm}   & $\Bphi^{\tilde 0 \check a}$ & $\frac{1}{\sqrt 2}\Sigma^2 \vert\gamma_{\check a}\vert$ & singlet 
& $\gamma_{\check a}$ \\
\rule[-.3cm]{0cm}{.8cm}  & $\Bphi^{\alpha \check a}_{1,2}$ & $\frac{1}{\sqrt 2}\Sigma^2  \vert \gamma \mp \gamma_{\check a}   \vert$ &
singlet& $\gamma \mp \gamma_{\check a}$\\
\hline
\end{tabular}
\caption{Decomposition of the $\cN = 4$ fields. $c_{\rm field} = \pm 1$ determines the SO(4) representation for 
the massive fields, see \eqref{cBB_choice} for $c_{\BB}$, \eqref{cBpsi_choice} for $c_{\Bpsi}$, and \eqref{cBlambda_choice} for $c_{\Blambda}$.}
\label{field_decomposition}
\end{table}

\setlength\extrarowheight{5pt}
\begin{table}
 \centering
\begin{tabular}{|l||l|l|l|}
\hline
\rule[-.3cm]{0cm}{.8cm}  multiplets & fields &  mass & charge\\
\hline\hline
\rule[-.3cm]{0cm}{.8cm}  1 gravity & $g_{\mu\nu},A^{\tilde 0}_\mu,\psi_\mu^\alpha$ 
 & 0 & 0\\
\hline
\rule[-.3cm]{0cm}{.8cm}  1 gravitino & $\Bpsi_\mu, \BB_{\mu\nu\,\alpha}$  & $\frac{1}{\sqrt 2} \Sigma^2  \gamma$ & $\gamma$\\
\hline
\multirow{2}{*}{$(1+ n -2n_T )$ vector} & \rule[-.2cm]{0cm}{.7cm}  $A^0_\mu,\chi_\alpha,\Sigma$  & 0 & 0\\
\cline{2-4}
& \rule[-.3cm]{0cm}{.8cm}  $A_\mu^{\tilde a},\lambda^{\tilde a}_\alpha,\phi^{\tilde 0 \tilde a}$ 
&0 &0\\
\hline
\rule[-.3cm]{0cm}{.8cm}  $n_T$ tensor & $\BB_{\mu\nu}^{\check a},\Blambda^{\check a}_\alpha,\Bphi^{\tilde 0 \check a}$ 
 & $\frac{1}{\sqrt 2}\Sigma^2 \vert \gamma_{\check a}\vert$ & $ \gamma_{\check a}$\\
\hline
\multirow{2}{*}{$n$ hyper} & \rule[-.3cm]{0cm}{.8cm} $\Blambda^{\tilde a},\Bphi^{\alpha\tilde a}$  &$\frac{1}{\sqrt 2} \Sigma^2  \gamma$ & $\gamma$ \\
\cline{2-4}
& \rule[-.3cm]{0cm}{.8cm}  $\Blambda^{\check a}_{1,2},\Bphi^{\alpha \check a}_{1,2}$ 
 & $\frac{1}{\sqrt 2}\Sigma^2  \vert \gamma \mp \gamma_{\check a}  \vert$ & $  \gamma \mp \gamma_{\check a}  $\\
\hline
\end{tabular}
\caption{$\cN = 2$ multiplets in the vacuum}
\label{vacuum_multiplets}
\end{table}

\subsection{General $\cN =2$ action and classical matching} \label{Classicaltruncation}

We are now in the position to derive the classical $\cN= 2$ effective action 
for the massless modes. In order to do that we simply truncate the $\cN=4$ 
action to the massless sector. The discussion of the quantum corrections can be found in 
the next subsection.

To begin with we recall the canonical form of 
a general $\cN=2$ ungauged supergravity theory.
The dynamics of the gravity-vector sector is entirely specified 
in terms of a cubic potential
\beq \label{canonical-cN}
\mathcal{N} = \tfrac{1}{3!} k_{I J K} M^{I} M^{J} M^{K}\, ,
\eeq
where $M^I, 1,\dots, n-2 n_T +2$ are very special real coordinates and $k_{IJK}$ is a 
symmetric tensor. The $M^I$ naturally combine with the vectors $A^I$ of the 
theory. However, since the vector in the gravity multiplet is not accompanied 
by a scalar degree of freedom, the $M^I$ have to satisfy one constraint.  
In fact, the $\cN=2$ scalar field space is identified with the hypersurface 
\begin{equation} \label{very-special-geometry-constraint}
\mathcal{N} \overset{!}{=} 1 \;.
\end{equation}
The gauge coupling function and the metric are then obtained as
\begin{equation} \label{gauge-coupling-function}
 G_{I J} = \left[ -\tfrac{1}{2} \partial_{M^I} \partial_{M^J} \log \mathcal{N} \right]_{\mathcal{N}=1} \ .
\end{equation}
The bosonic two-derivative Lagrangian is then given by
\begin{align} \label{5d_action_canonical}
\cL_{\rm can} =  &-\tfrac{1}{2} R 
 - \tfrac{1}{2} G_{ I J} \partial_\mu M^{I} \partial^\mu M^{J} -\tfrac{1}{4} G_{I J} F^{I}_{\mu \nu} F^{\mu \nu \, J}
\nn \\
&
 + \tfrac{1}{48} \epsilon^{\mu \nu \rho \sigma\lambda} k_{I J K} A^{I}_\mu  F^{J}_{\nu \rho} F^{K}_{\sigma \lambda} -H_{u v} \partial_\mu h^u \partial^\mu h^{v} \;.
\end{align}
Here we included the kinetic term for the hypermultiplet scalars $h^u$ with metric $H_{uv}$.

The canonical Lagrangian \eqref{5d_action_canonical} has to be compared 
with the truncated $\cN=4$ theory. In our set-up we found the vectors $(A^I) = (A^0,\BA^{\tilde 0},A^{\tilde a})$, which 
sets the index range for $I$.
The \emph{massless} scalars in the effective theory (except for $\Sigma$) are most conveniently described by 
$SO(5,n)$-rotated elements of the coset space
\begin{align}\label{normalized_Vs}
 \hat\cV := \langle \cV \rangle^{-1} \cV = \text{exp} \big( \phi^{ma} [t_{ma}]\big)
\end{align}
This is in contrast to the analysis of  the \emph{massive} scalar spectrum, for which 
it is efficient to consider the fluctuations $\phi^{ma}$ as it was done in the last section.
Restricting to the $\cN =2$ vector multiplets and
truncating the massive modes
$\phi^{\tilde 0 \hat a}$ and $\phi^{\hat m \tilde a}$, the only
remaining elements of the coset space are 
\begin{align} \label{Vhats}
 \tensor{\hat\cV}{_{\tilde 0}^{\tilde 0}}\, , \quad \tensor{\hat\cV}{_{\tilde 0}^{\tilde a}}\, , \quad 
 \tensor{\hat\cV}{_{\tilde a}^{\tilde 0}}\, , \quad \tensor{\hat\cV}{_{\tilde a}^{\tilde b}}\, .
\end{align}
In fact, it turns out that all couplings involving the elements \eqref{Vhats} can be 
expressed as functions of $\tensor{\hat\cV}{_{\tilde 0}^{\tilde a}}$ alone. 
In order to do that, one uses the relations
\begin{align}
\tensor{\hat\cV}{_{\tilde 0}^{\tilde 0}} = 
\sqrt{1+ \tensor{\hat\cV}{_{\tilde 0}^{\tilde a}}\tensor{\hat\cV}{_{\tilde 0}_\, _{\tilde a}}}\, , \qquad
 \tensor{\hat\cV}{_{\tilde a}^{\tilde 0}}=\tensor{\hat\cV}{_{\tilde 0}^{\tilde a}}\, , \qquad
 \tensor{\hat\cV}{_{\tilde a}^{\tilde c}}\tensor{\hat\cV}{_{\tilde b}_\, _{\tilde c}} = \delta_{\tilde a \tilde b}
 + \tensor{\hat\cV}{_{\tilde a}^{\tilde 0}} \tensor{\hat\cV}{_{\tilde b}^{\tilde 0}}\, ,  \qquad
 \tensor{\hat\cV}{_{\tilde a}^{\tilde b}}\tensor{\hat\cV}{_{\tilde 0}_\, _{\tilde b}} =
 \tensor{\hat\cV}{_{\tilde 0}^{\tilde 0}} \tensor{\hat\cV}{_{\tilde a}^{\tilde 0}} \, . 
\end{align}
The element  $\tensor{\hat\cV}{_{\tilde 0}^{\tilde a}}$ itself can be expanded as 
\beq
   \tensor{\hat\cV}{_{\tilde 0}^{\tilde a}}= \text{exp} \big( \phi^{\tilde 0 \tilde a} [t_{\tilde 0 \tilde a}]\big){_{\tilde 0}^{\ \, \tilde a}} \, ,
\eeq 
after truncating all massive modes.
This implies, in particular, that $\tensor{\hat\cV}{_{\tilde 0}^{\tilde a}}$ has no dependence on $\phi^{\hat m \hat a}$. 
Therefore, the effective action of the scalars in the $\cN=2$ vector multiplets decouples from the potentially massless
scalars in the hypermultiplets as expected from $\cN=2$ supersymmetry.

The reduced Lagrangian then takes the simple form
\begin{align}\label{eff_action}
 e^{-1} \cL_{\textrm{class}}=&
 - \frac{1}{2} R - H_{\Lambda\Sigma}^{pq}\, \partial_\mu h^\Lambda_p \, \partial^\mu h^\Sigma_q  
 - \frac{3}{2} \Sigma^{-2}\, \partial_\mu \Sigma \, \partial^\mu \Sigma \nn \\
& -\frac{1}{2} \Big ( \delta_{\tilde a \tilde b} - 
 \frac{1}{1 + \tensor{\hat\cV}{_{\tilde 0}^{\tilde c}}\, \tensor{\hat\cV}{_{\tilde 0}_\, _{\tilde c}}}
 \tensor{\hat\cV}{_{\tilde 0}_\, _{\tilde a}}\tensor{\hat\cV}{_{\tilde 0}_\, _{\tilde b}}\Big )\,
 \partial_\mu \tensor{\hat\cV}{_{\tilde 0}^{\tilde a}} \, \partial^\mu \tensor{\hat\cV}{_{\tilde 0}^{\tilde b}}
  \nn \\
  &- \frac{1}{4} \Sigma^{-4}\, F^{0}_{\mu\nu} F^{\mu\nu\,0} + \Sigma^2 
   \sqrt{1+\tensor{\hat\cV}{_{\tilde 0}^{\tilde b}}\tensor{\hat\cV}{_{\tilde 0}_\, _{\tilde b}}}\,\tensor{\hat\cV}{_{\tilde 0}_\, _{\tilde a}}\,
   F^{\tilde 0}_{\mu\nu} F^{\mu\nu\,\tilde a} \nn \\ 
 &- \frac{1}{4} \Sigma^2 \Big ( 3 + 2\, \tensor{\hat\cV}{_{\tilde 0}^{\tilde a}}\tensor{\hat\cV}{_{\tilde 0}_\, _{\tilde a}}  \Big )\,
 F^{\tilde 0}_{\mu\nu} F^{\mu\nu\,\tilde 0}
 - \frac{1}{4} \Sigma^2 \Big ( \delta_{\tilde a \tilde b } + 
 2\, \tensor{\hat\cV}{_{\tilde 0}_\, _{\tilde a}} \tensor{\hat\cV}{_{\tilde 0}_\, _{\tilde b}} \Big )\,
 F^{\tilde a}_{\mu\nu} F^{\mu\nu\,\tilde b} \nn \\
 & +\frac{1}{4\sqrt 2} \epsilon^{\mu\nu\rho\sigma\tau} A_\mu^0 F^{\tilde 0}_{\nu\rho} F^{\tilde 0}_{\sigma\tau}  
 - \frac{1}{4\sqrt 2} \epsilon^{\mu\nu\rho\sigma\tau} A_\mu^0 F^{\tilde a}_{\nu\rho} F^{\tilde a}_{\sigma\tau} \, ,
\end{align}
where $H_{\Lambda\Sigma}^{pq}$ is the metric of the quaternionic manifold parametrized by the scalars in the massless hypermultiplets that
we, however, do not discuss any further in this paper.
Therefore, by comparison of \eqref{eff_action} with \eqref{5d_action_canonical} we find the identifications
\begin{align} \label{def-Ms}
 M^0 = \frac{1}{\sqrt 2} \Sigma^2\, , \qquad M^{\tilde 0} = \Sigma^{-1} \tensor{\hat\cV}{_{\tilde 0}^{\tilde 0}}\, , \qquad
 M^{\tilde a} = \Sigma^{-1} \tensor{\hat\cV}{_{\tilde 0}^{\tilde a}} \, ,
\end{align}
and the real potential
\begin{align} \label{Nclass}
 \cN = \frac{1}{2} k_{0 \tilde 0 \tilde 0} M^0 M^{\tilde 0} M^{\tilde 0}
 + \frac{1}{2} k_{0 \tilde a \tilde a} M^0 M^{\tilde a} M^{\tilde a}
 = \sqrt 2 M^0 M^{\tilde 0} M^{\tilde 0} - \sqrt 2 M^0 M^{\tilde a} M^{\tilde a} \, .
\end{align}
This result specifies the constant tensors $k_{IJK}$ at the classical level. 
It is interesting to realize that the constraint $\cN \overset{!}{=}1$ translates with the identifications \eqref{def-Ms}
into the condition \eqref{eta_viaV} for the elements of the coset space.
We conclude that the very special real manifold is the 
coset space 
\beq
    SO(1,1) \times \frac{SO(1,n-2n_T)}{SO(n-2n_T)}\ ,
\eeq
which is the subspace of \eqref{coset_def} spanned by the 
massless scalars in the vector multiplets. 

\subsection{One-loop effects and Chern-Simons terms} \label{oneloopeffects}

In this section we determine the one-loop corrections to the 
gravity-vector sector of the $\cN=2$ theory specified in \autoref{Classicaltruncation}.
We focus on this sector, since the corrections due to integrating 
out massive fields are independent of the supersymmetry breaking 
scale and masses of the fields running in the loop. Let us stress 
that due to the preserved $\cN=2$ supersymmetry and the fact that the Chern-Simons 
terms can only receive constant corrections, the integrating
out process can only perturbatively correct the gravity-vector sector at 
the one-loop level. 

To obtain the one-loop corrected $\cN$ an analysis 
of the Chern-Simons terms is sufficient. The 
explicit loop computations were performed in \cite{Bonetti:2012fn,Bonetti:2013ela} 
and we can simply apply these results to our set-up. 
The studied Chern-Simons terms are of the form
\begin{align} \label{CS-terms}
 e^{-1} \cL_{CS} =  \frac{1}{48} \epsilon^{\mu\nu\rho\sigma\tau} 
 k_{I J K} A^I_\mu F^J_{\nu\rho} F^K_{\sigma\tau} 
 + \frac{1}{48} \epsilon^{\mu\nu\rho\sigma\tau} 
 k_{I} A^I_\mu \tensor{R}{^a_b_\nu_\rho} \tensor{R}{^b_a_\sigma_\tau} \, . 
\end{align}
Note that in \cite{Bonetti:2013ela} also the one-loop corrections to the gauge-gravitational Chern-Simons 
term, the second term in \eqref{CS-terms}, were computed and we include the result for completeness. 
The one-loop corrections arise from integrating out massive 
fields that are charged under some gauge fields $A^I$. 
We denote the charges of massive gravitini, tensors, and spin-1/2 fermions 
collectively by $q_I$ and denote by $c_{\Bpsi},c_\BB$ and $c_{\Blambda}$ the $\pm 1$ choice of 
representation under the massive little group $SO(4)$.
The one-loop terms are calculated according to the \autoref{one-looptable}.
These results hold when integrating out massive tensors, gravitini and spin-1/2 fermions
with actions \eqref{standard_complex_tensor}, \eqref{e:lagrangian_gravitino}, \eqref{e:lagrangian_1/2}
and are thus readily applied to our supersymmetry breaking set-up. 
\begin{table}[h!]
\begin{center}
\begin{tabular}{|rccc|}
\hline
 \rule[-.3cm]{0cm}{.7cm}  & spin 1/2 & tensor & spin 3/2\\
 \hline
\rule[-.3cm]{0cm}{.7cm} $k_{I JK}$ = & $-q_{I}q_{J}q_{K} \cdot c_{\Blambda}$ & $-q_{I}q_{J}q_{K} \cdot (-4 c_{\BB})$ 
& $-q_{I}q_{J}q_{K} \cdot (5 c_{\Bpsi})$\\
\rule[-.3cm]{0cm}{.7cm} 
$k_{I}$ = & $-\frac{1}{8} q_{I}\cdot c_{\Blambda}$ & $-\frac{1}{8} q_{I}\cdot (8 c_\BB )$ & $- \frac{1}{8} q_{I}\cdot (-19 c_{\Bpsi})$\\
\hline
\end{tabular}
\caption{One-loop corrections due to integrating out massive spin-1/2 fermions, tensors, and spin-3/2 fermions.}
\label{one-looptable}
\end{center}
\end{table}

Since all massive fields are only charged under $A^0$, the classical terms in 
\eqref{Nclass} are unmodified. The fully quantum corrected result therefore reads
\begin{align}
k_{0\tilde 0 \tilde 0} =
 -k_{0\tilde a \tilde a}= 2 \sqrt 2 \, .
\end{align}
Using \autoref{one-looptable} we find that the massive states summarized in \autoref{field_decomposition} induce by the 
one-loop couplings
\begin{align} \label{one-loop_k000}
 k_{000} = -\Big [ (-3- n + 2n_T) \gamma^3 -2 \sum_{\check a} \vert  \gamma_{\check a} \vert^3
 + \sum_{\check a}  \vert  \gamma -  \gamma_{\check a} \vert^3 + \sum_{\check a}  \vert  \gamma  + \gamma_{\check a} \vert^3 \Big ] \, .
\end{align}
Furthermore, we find the gravitational one-loop Chern-Simons terms
\begin{align}\label{one-loop_k0}
 k_0 = - \frac{1}{8} \Big [ (-3 - n + 2n_T ) \gamma + 10 \sum_{\check a}  \vert \gamma_{\check a} \vert 
 + \sum_{\check a}  \vert \gamma - \gamma_{\check a} \vert + \sum_{\check a}  \vert \gamma + \gamma_{\check a} \vert \Big ] \, . 
\end{align}

The existence of new Chern-Simons couplings implies that the effective theory still sees remnants of the underlying $\cN=4$ theory at 
arbitrarily low energy scales. In fact, the mass of the fields listed in \autoref{vacuum_multiplets} 
can be made arbitrarily large by choosing the VEV of the modulus $\Sigma$. The 
 constants $\gamma$ and $\gamma_{\check a}$ appearing in \eqref{one-loop_k000}  and \eqref{one-loop_k0} are the imaginary parts 
 of the eigenvalues of $\xi^{MN}$ and therefore independent of the VEVs of the fields. 

\section{Conclusions}

In this work we have studied partial supersymmetry breaking in 
$\cN=4$ supergravity theories. We focused on a 
certain subclass of gauged supergravity theories that 
possess tensor fields with first order kinetic terms.
These are characterized by a non-trivial embedding 
tensor $\xi_{MN}$. The presence of these couplings 
also ensures that they can become massive by a tensorial 
Higgs mechanism and are gauged by a distinctive $U(1)$ 
vector field $A^0$. 
The Higgs mechanism allows that the tensors  
acquire three degrees of freedom by absorbing a 
dynamical vector. Clearly, this is just 
a special case of the mechanisms that are employed to 
construct general supergravity theories using 
tensor hierarchies \cite{Samtleben:2008pe,Hartong:2009vc}. We show that 
the presence of these couplings permits the existence of 
simple $\cN=2$ supersymmetric Minkowski vacua.

The fluctuations around the $\cN=2$ supersymmetric 
Minkowski vacua are split into massive and massless 
$\cN=2$ multiplets. In fact, starting with the $\cN=4$
gravity multiplet and $n$ vector multiplets, we showed that 
the massless $\cN=2$ spectrum consists of the gravity multiplet, 
$n-2n_T+1$ vector multiplets, and $n_H$ hypermultiplets. In addition 
one finds a massive spin-3/2 multiplet, $n_T$ complex 
massive tensor multiplets, and $n-n_H$ massive hypermultiplets. 
The massive spin-3/2 multiplet contains in addition to the gravitino 
degrees of freedom also two complex massive tensors. 
The number $n_T$ is given by $2n_T = \text{rank}( \xi_{MN} )- 4$,
where the $4$ corresponds to the massive tensors in the spin-3/2 multiplet. 
The degeneracy $n_H$ is given by the number of hypermultiplets
that satisfy the masslessness condition \eqref{massless_condition}.

We also determined the $\cN=2$ low-energy effective action
for the massless modes particularly focusing on the vector multiplets. 
In order to do that we first extracted the classical $\cN=2$ couplings 
obtained by dropping all massive fields. We argued, however, that 
it is crucial to evaluate the 
actions for the massive tensors, and massive spin-3/2 and 
spin-1/2 fermions that are charged under the 
vector field $A^0$. This is due to the fact, that these massive 
fields induce corrections to the kinetic terms and Chern-Simons 
terms of $A^0$ at one-loop. Crucially, these corrections are independent of 
the supersymmetry breaking scale and thus have to be included despite the 
fact that they are of loop order. In other words, from the Chern-Simons terms 
of the effective theory, one can extract information about the underlying $\cN=4$
theory. This is of similar spirit as the discussion in \cite{Bonetti:2013cza}, where five-dimensional one-loop 
Chern-Simons terms were used to extract information about an underling 
six-dimensional theory. 

Let us briefly discuss possible applications of our results and comment of 
further extensions. Five-dimensional theories with $\cN=4$ supersymmetry 
can arise, for example, in compactifications of M-theory on $K3 \times T^2$, or 
Type IIB supergravity on $K3 \times S^1$. In order to obtain a gauged 
$\cN=4$ theory the $T^2$ or $S^1$ reduction can, for example, be made non-trivial 
by demanding Scherk-Schwarz boundary conditions. In fact, in such reductions
one expects that certain states are charged under the Kaluza-Klein vector that 
is identified with our $A^0$. In particular, it would be interesting to 
carry out a Scherk-Schwarz reduction of a six-dimensional
$(2,0)$ theory on a circle and determine its $\cN=2$ vacua. 

More involved five-dimensional gauged $\cN=4$ supergravity theories 
arise from M-theory on a general SU(2) structure manifold, or from 
Type IIB on a squashed Sasaki-Einstein 
manifold \cite{Gauntlett:2004zh,Cassani:2010uw,Gauntlett:2010vu,Liu:2010sa,Cassani:2010na,Bena:2010pr,Bah:2010cu,OColgain:2011ng}.  
These theories in general also admit non-Abelian gaugings and allow 
for non-trivial Anti-de Sitter vacua. In a future project 
we hope to extend our analysis to include these more 
involved situations. We expect that similar to our 
discussion for the Minkowski vacua also one-loop effects will be essential 
when evaluating the effective theory.

\subsubsection*{Acknowledgments}

We would like to thank Federico Bonetti, Stefan Hohenegger, Jan Keitel, Jan Louis,
Severin L\"ust, Tom Pugh, and Hagen Triendl for interesting discussions and comments.
This work was supported by a research grant of the Max Planck Society.

\appendix
\section{Conventions and identities} \label{App-Conventions}
\subsection{Spacetime conventions}
In this section we state the conventions of Riemannian geometry adopted in this paper.
We denote curved five-dimensional spacetime indices by Greek letters $\mu,\nu,\dots$. Antisymmetrizations
of any kind of indices are made with weight one, i.e.~include a factor of $1/n!\,$.
For the five-dimensional metric $g_{\mu\nu}$ we use the $(-,+,+,+,+)$ sign convention and we choose a negative sign in front of the Einstein-Hilbert term.
Furthermore we set
\begin{align}
 \kappa^2  = 1 \, .
\end{align}
We adopt the following convention for the Levi-Civita tensor with curved indices $\epsilon_{\mu\nu\rho\lambda\sigma}$, 
$\epsilon^{\mu\nu\rho\lambda\sigma}$
\begin{align}
 \epsilon_{01234} = + e \, , \qquad  \, , \epsilon^{01234} = - e^{-1}\, ,
\end{align}
where $e = - \sqrt{-\det g_{\mu\nu} }\,$.

We denote the five-dimensional spacetime gamma matrices by $\gamma_\mu$, satisfying
\begin{align}
 \lbrace \gamma_\mu , \gamma_\nu \rbrace = 2 g_{\mu\nu} \, .
\end{align}
Antisymmetrized products of gamma matrices are defined as
\begin{align}
 \gamma_{\mu_1 , \dots , \mu_k } := \gamma_{[ \mu_1}\gamma_{\mu_2}\dots \gamma_{\mu_k ]}\, . 
\end{align}
The charge conjugation matrix $C$ is chosen such that
\begin{align}
 C^T = -C = C^{-1}
\end{align}
and it fulfills the relation
\begin{align}
 C \gamma_\mu C^{-1} = ( \gamma_\mu )^T \, .
\end{align}
All massless spinors in this paper are symplectic Majorana. In the $\cN=4$ theory they are subject to the condition
\begin{align}
 \bar \chi^i := (\chi_i)^\dagger \gamma_0 = \Omega^{ij} \chi_j^T C \, ,
\end{align}
where $i,j =1, \dots , 4$ and $\Omega^{ij}$ is the $USp(4)$ symplectic form defined in \eqref{e:properties_omega}.
The $\cN=2$ symplectic Majorana condition reads
\begin{align}
 \bar \chi^\alpha := (\chi_\alpha)^\dagger \gamma_0 = \varepsilon^{\alpha\beta} \chi_\beta^T C \, ,
\end{align}
where $\alpha,\beta = 1,2$ and $\varepsilon^{\alpha\beta}$ is the two-dimensional epsilon tensor.
Let us state a set of useful relations between bilinears of symplectic Majorana spinors
\begin{align}
 \bar \chi^i \gamma_{\mu_1 \dots \mu_k} \lambda^j =
 \begin{cases}
  + \bar \lambda^j \gamma_{\mu_1 \dots \mu_k} \chi^i \quad (k= 0,1,4,5) \\
  - \bar \lambda^j \gamma_{\mu_1 \dots \mu_k} \chi^i \quad (k= 2,3)
 \end{cases}
\end{align}
and
\begin{align}
 \bar \chi^i \gamma_{\mu_1 \dots \mu_k} \lambda_i =
 \begin{cases}
  - \bar \lambda^i \gamma_{\mu_1 \dots \mu_k} \chi_i \quad (k= 0,1,4,5) \\
  + \bar \lambda^i \gamma_{\mu_1 \dots \mu_k} \chi_i \quad (k= 2,3)\, .
 \end{cases}
\end{align}
Both relations are are also true for symplectic Majorana spinors carrying indices $\alpha,\beta = 1,2\,$. 

\subsection{SO(5) gamma matrices}
The scalars in the vector multiplets of the $\cN = 4$ theory span the coset manifold
\begin{align}
 \cM_{5,n} = \frac{SO(5,n)}{SO(5)\times SO(n)}\, .
\end{align}
The coset representatives are denoted by $(\tensor{\cV}{_M^m} , \tensor{\cV}{_M^a})$ and their properties are discussed
in \autoref{N=4Gen}.
In this section we collect some properties of the $SO(5)$ gamma matrices $\tensor{\Gamma}{_m^i^j}$, which satisfy
\begin{align}
 \lbrace \Gamma_m , \Gamma_n \rbrace = 2 \delta_{mn} \id_4 \, .
\end{align}
Furthermore we have
\begin{align}
 \tensor{\Gamma}{_m^i^j} = - \tensor{\Gamma}{_m^j^i}\, , \qquad \tensor{\Gamma}{_m_\, _i^i}  = 0 \, , 
 \qquad (\tensor{\Gamma}{_m^i^j})^* = \Omega_{ik} \Omega_{jl} \tensor{\Gamma}{_m^k^l}
\end{align}
as well as
\begin{align}
 \tensor{\Gamma}{^m^\, ^i^j}\tensor{\Gamma}{_m_\, _k_l} = 4 \delta^{ij}_{kl} - \Omega^{ij} \Omega_{kl} \, .
\end{align}

Note that one can use $\Gamma_{mn} := \Gamma_{[m}\Gamma_{n]}$ in order to switch between antisymmetric tensors
$T_{mn}$ and symmetric tensors $T_{ij}$ in the following way
\begin{align}\label{tensor_correspondence}
 T_{ij} = T_{mn} \tensor{\Gamma}{^m^n_i_j} \, , \qquad T_{mn} = \frac{1}{8}\Gamma_{mn\, ij} T^{ij}\, ,
\end{align}
which can be checked using the identity
\begin{align}
 \Gamma_{mn\, ij}\Gamma^{pq\, ij} = 8 \delta^{pq}_{mn}\, .
\end{align}
This correspondence is for example used in \eqref{ferm_mass}.

\section{Derivation of the mass terms and couplings} \label{mass_appendix}

In this section we explicitly derive the masses for the spin-1/2 fermions and scalars 
induced by the supergravity breaking. We also evaluate the charges of the spin-1/2 fermions under the Abelian gauge field $A^0$. 

We shortly note that the Lagrangian
of a massive spin-1/2 Dirac spinor is given by
\begin{align}
\label{e:lagrangian_1/2}
 e^{-1} \cL_\Blambda = -\bar \Blambda \slashed \cD \Blambda +  c_{\Blambda}  m \bar \Blambda \Blambda
 \, , \qquad c_{\Blambda} =\pm 1 \, .
\end{align}
The $c_{1/2}$ in \eqref{e:lagrangian_1/2} refers to the two inequivalent spinor representations of the massive little group, 
in detail
\begin{align}  \label{cBlambda_choice}
 c_{\Blambda} &= + 1 \Leftrightarrow (\frac{1}{2}, 0 )\textrm{ of } SO(4) \cong  SU(2) \times SU(2) \\
 c_{\Blambda} &= - 1 \Leftrightarrow (0 , \frac{1}{2})\textrm{ of } SO(4) \cong SU(2) \times SU(2) \, . \nn
\end{align}
The mass of the physical mode is given by $m$.

\subsection{Fermion masses}
Let us now investigate the masses the $\cN = 4$ gaugini acquire from supersymmetry breaking.
The mass terms read
\begin{align}
\label{e:fermion_quadratic}
 e^{-1}\cL_{\lambda , \, \textrm{mass}} = &i \Big ( \frac{1}{2 \sqrt 2} \Sigma^2 \xi_{ab}\delta^j_i
 - \frac{3}{2\sqrt 6}\tensor{A}{_1_\, _i^j}\delta_{ab}\Big ) \bar \lambda^{ia}\lambda^b_j = \nn \\
 = & 0 \cdot \bar \lambda^{\alpha \tilde a}\lambda^{\tilde b}_\alpha
  + \frac{1}{2 \sqrt 2} i \Sigma^2 \xi_{\hat a \hat b} 
 \bar \lambda^{\alpha \hat a}\lambda^{\hat b}_\alpha
  - \frac{3}{2\sqrt 6}i \tensor{A}{_1 _\, _{\dot\alpha}^{\dot\beta}}\delta_{\tilde a \tilde b}\bar 
  \lambda^{\dot\alpha \tilde a}\lambda^{\tilde b}_{\dot\beta} \nn \\
 &i \Big ( \frac{1}{2 \sqrt 2} \Sigma^2 \xi_{\hat a \hat b} \delta^{\dot\beta}_{\dot\alpha}
 - \frac{3}{2\sqrt 6}\tensor{A}{_1 _\, _{\dot\alpha}^{\dot\beta}}
 \delta_{\hat a \hat b}\Big ) \bar \lambda^{\dot\alpha \hat a}\lambda^{\hat b}_{\dot\beta}\, .
 \end{align}
 Let us discuss the four different types of fields separately.
 
\begin{enumerate} 

\item $\lambda^{\tilde a}_\alpha$ \\
We observe that the $\lambda^{\tilde a}_\alpha$ stay massless.
Thus we have $2(n-2n_T)$ massless spin-1/2 fermions supplemented by a symplectic Majorana condition.

\item $\lambda^{\hat a}_\alpha$ \\
For the fermions $\lambda^{\hat a}_\alpha $.
We write the mass terms using the split \eqref{split_lambdas} as
\begin{align}
 \frac{1}{2 \sqrt 2} i \Sigma^2 \xi_{\hat a \hat b}\bar\lambda^{\alpha \hat a}\lambda^{\hat b}_{\alpha}&=
  \frac{1}{2 \sqrt 2} i \Sigma^2 \sum_{\check a}\gamma_{\check a}\varepsilon_{kl}
 \bar \lambda^{\alpha [\check a k]}\lambda^{[\check a l]}_{\alpha} \nn \\
 & = \frac{1}{\sqrt 2}\Sigma^2 \sum_{\check a}\gamma_{\check a} 
 \bar\Blambda^{\alpha\check a}\Blambda_\alpha^{\check a}\, ,
\end{align}
with $m,n$ both taking values $1,2$.
Here we redefined the fermions by introducing $\Blambda^{\check a}_\alpha$ as in \eqref{compl_fermions_1}
and drop the symplectic Majorana condition, such that the mass terms become diagonal.
Let us now have a look how the corresponding kinetic terms transform under this redefinition
\begin{align}
 -\frac{1}{2}\sum_{\check a}\big( \bar\lambda^{\alpha [\check a 1]}\slashed \partial \tensor{\lambda}{_\alpha^{[\check a1]}}
 + \bar\lambda^{\alpha [\check a 2]}\slashed \partial \tensor{\lambda}{_\alpha^{[\check a 2]}}\big)=
 -\sum_{\check a}\bar{\Blambda}^{\alpha \check a }\slashed \partial \Blambda_\alpha^{\check a}\, .
\end{align}
The computation of the charge under $A^0$ proceeds as for the mass terms. By comparing the 
action with \eqref{e:lagrangian_1/2} we find
\begin{align}
 m_{\Blambda^{\check a}_\alpha} = \frac{1}{\sqrt 2} \Sigma^2  \vert \gamma_{\check a}\vert\, , \qquad
 c_{\Blambda^{\check a}_\alpha} =  \sign \gamma_\alpha \, , \qquad 
 q_{\Blambda^{\check a}_\alpha} =  \gamma_{\check a}\, .
\end{align}

\item $\lambda^{\tilde a}_{\dot \alpha}$\\
The structure of mass terms of the fermions $\lambda^{\tilde a}_{\dot \alpha}$
is similar to those of the gravitino masses. In particular, the diagonalization 
procedure of the gravitino mass terms automatically
diagonalizes the mass terms of the $\lambda^{\tilde a}_{\dot\alpha }$. Again we move from symplectic Majorana spinors to
Dirac spinors $\Blambda^{\tilde a}$ using \eqref{compl_fermions_2}.
We find
\begin{align}
 m_{\Blambda^{\tilde a}} = \frac{1}{\sqrt 2}  \Sigma^2 \gamma \, , \qquad
 c_{\Blambda^{\tilde a}} = -1 \, ,\qquad  q_{\Blambda^{\tilde a}} =  \gamma \, .
\end{align}
The $\Blambda^{\tilde a}$ are in the $(0, \frac{1}{2})$ representation of $SO(4)$.

\item $\lambda^{\hat a}_{\dot\alpha }$\\
The mass terms become after the split \eqref{split_lambdas}
\begin{align}
 i \sum_{\check a} \Big ( \frac{1}{2 \sqrt 2} \Sigma^2 \gamma_{\check a} \varepsilon_{kl} \delta^{\dot\beta}_{\dot\alpha}
 - \frac{3}{2 \sqrt 6}\tensor{A}{_1 _\, _{\dot\alpha}^{\dot\beta}}
 \delta_{kl}\Big ) \bar \lambda^{\dot\alpha [\check a k]}\lambda^{[\check a l]}_{\dot\beta}\, .
\end{align}
 We redefine and use Dirac spinors $\Blambda^{\check a}_1$ and $\Blambda^{\check a}_2$ given in \eqref{compl_fermions_2}. 
The mass terms then become
\begin{align}
  \sum_{\check a} \Big [ \frac{1}{\sqrt 2} \Sigma^2 \gamma_{\check a} 
 ( \bar{\Blambda}^{\check a}_1 \Blambda_1^{\check a} - 
 \bar{\Blambda}^{\check a}_2 \Blambda_2^{\check a} ) -
 \frac{1}{\sqrt 2} \Sigma^2  \gamma ( \bar{\Blambda}^{\check a}_1 \Blambda_1^{\check a} +  
 \bar{\Blambda}^{\check a}_2 \Blambda_2^{\check a} ) \Big ]\, .
\end{align}
The kinetic terms are unaffected.
We conclude that
\begin{align}
 &m_{\Blambda^{\check a}_1}= \frac{1}{\sqrt 2}\Sigma^2  \vert \gamma_{\check a} - \gamma   \vert \, ,& \qquad 
 &m_{\Blambda^{\check a}_2}= \frac{1}{\sqrt 2}\Sigma^2  \vert - \gamma_{\check a} - \gamma  \vert \, ,& \\
 &c_{\Blambda^{\check a}_1}= \sign (\gamma_{\check a} - \gamma  ) \, , & \qquad 
 &c_{\Blambda^{\check a}_2}= \sign (- \gamma_{\check a} - \gamma ) \, , &\\
 &q_{\Blambda^{\check a}_1}= \gamma_{\check a} - \gamma   \, ,& \qquad 
 &q_{\Blambda^{\check a}_2}= - \gamma_{\check a} - \gamma  \, .&
\end{align}

\end{enumerate}

\subsection{Scalar masses}
Lastly we investigate the scalar degrees of freedom in the vacuum (except of $\Sigma$).
In order to derive the scalar masses we insert the expansion \eqref{cVexpansion}
into the scalar potential \eqref{scalar_potential} 
\begin{align}
 e^{-1} \cL_{\textrm{pot}} = - \frac{1}{16} \xi^{MN} \xi^{PQ} \Sigma^4
 &\Big [\langle\cV\rangle\exp \big (  \sum_{m,a} \phi^{ma} [t_{ma}]\Big ) 
 \exp \big (  \sum_{n,b} \phi^{nb} [t_{nb}]\big )^T \langle\cV \rangle^T\Big ]_{MP} \times \nn \\
 &\Big [\langle\cV\rangle \exp \big (  \sum_{p,c} \phi^{pc} [t_{pc}]\big ) 
 \exp \big (  \sum_{q,d} \phi^{qd} [t_{qd}]\big )^T \langle\cV \rangle^T \Big ]_{NQ} \, .
\end{align}
To read off the mass terms of the scalars, we focus on the terms quadratic in $\phi^{ma}$
\begin{align}
 e^{-1} \cL_{\phi, \, \textrm{mass}} = - \frac{1}{16} \Sigma^4 \phi^{ma} \phi^{nb}
 (8 \xi_{mn} \xi_{ab} + 4 \delta_{mn} \xi_{ac} \tensor{\xi}{_b^c} + 4 \delta_{ab} \xi_{mp} \tensor{\xi}{_n^p})\, .
\end{align}

According to the index split \eqref{e:split} the scalar fields arrange in four different groups: 
\begin{enumerate}
 \item $\phi^{\tilde 0 \tilde a}$ \\
 The mass terms for these fields vanish:
 \begin{align}
  e^{-1} \cL_{\phi, \, \textrm{mass}} = 0 \, .
 \end{align}
 Thus we find $n - 2 n_T$ massless real scalar fields $\phi^{\tilde 0 \tilde a}$.
 
 \item $\phi^{\hat m \tilde a}$ \\
 The mass terms of these modes receive one contribution from the gauging $\xi^{mn}$
 \begin{align}
  e^{-1} \cL_{\phi, \, \textrm{mass}} = - \frac{1}{4}\gamma^2 \Sigma^4 \phi^{\hat m \tilde a} \phi^{\hat m \tilde a} \, .
 \end{align}
 We can now complexify the scalars as in \eqref{compl_scalars_1}  into the $2(n - 2 n_T)$ massive complex scalars $\Bphi^{\alpha \tilde a}$
 with mass\footnote{We stress that the kinetic terms
  for the scalars here and in the following are always automatically canonically normalized, even after
  the field redefinitions carried out in this section. This one can check explicitly by inserting the expansion
  \eqref{cVexpansion} into the $\cN =4$ scalar kinetic terms.}
 \begin{align}
  m_{\Bphi^{\alpha \tilde a}} = \frac{1}{\sqrt 2}  \Sigma^2 \gamma \, .
 \end{align}
 
 \item $\phi^{\tilde 0 \hat a}$ \\
 There is now solely a mass contribution from the gaugings $\xi^{ab}$
 \begin{align}
  e^{-1} \cL_{\phi, \, \textrm{mass}} = -\frac{1}{4}\Sigma^4 \sum_{\hat a} \gamma_{\check a}^2 \phi^{\tilde 0 \hat a}
  \phi^{\tilde 0 \hat a } \, .
 \end{align}
  Using the definition \eqref{compl_scalars_1} 
  one identifies $n_T$ massive complex scalar fields $\Bphi^{\tilde 0 \check a}$ with mass
  \begin{align}
   m_{\Bphi^{\tilde 0 \check a}} = \frac{1}{\sqrt 2} \Sigma^2 \vert \gamma_{\check a} \vert \, .
  \end{align}

  \item $\phi^{\hat m \hat a}$ \\
  We now face mass contributions both from $\xi^{mn}$ and $\xi^{ab}$
  \begin{align}
    e^{-1} \cL_{\phi, \, \textrm{mass}} = - \frac{1}{16}\Sigma^4 \sum_{\check a , \alpha} 
    (8 \gamma \gamma_{\check a} \varepsilon_{\dot\alpha\dot\beta}\varepsilon_{kl}
    + 4 \gamma^2  \delta_{\dot\alpha\dot\beta}\delta_{kl}
    + 4 \gamma_{\check a}^2  \delta_{\dot\alpha\dot\beta}\delta_{kl} )
    \phi^{[\alpha\dot\alpha ][\check a k]} \phi^{[\alpha\dot\beta ][\check a l]} \, .
  \end{align}
  One can check that the mass terms are diagonalized by the redefinitions 
  \eqref{compl_scalars_2} and \eqref{compl_scalars_3} to $4 n_T$ complex scalars
  $\Bphi^{\alpha \check a}_1$ and 
   $\Bphi^{\alpha \check a}_2$ with masses
  \begin{align}
   m_{\Bphi^{\alpha \check a}_1} = \frac{1}{\sqrt 2} \Sigma^2 \vert \gamma - \gamma_{\check a} \vert\, ,  \qquad
   m_{\Bphi^{\alpha \check a}_2 } = \frac{1}{\sqrt 2} \Sigma^2 \vert \gamma + \gamma_{\check a} \vert  \, .
  \end{align}
\end{enumerate}

\bibliography{Andreas,Thomas}
\bibliographystyle{utcaps}
\end{document}